\documentclass[11pt]{article}
\pdfoutput=1
\usepackage{jheppub}
\usepackage{amstext,amsmath,amssymb,amsfonts,bbm,amsthm}
\usepackage{physics}
\usepackage[latin1]{inputenc}
\usepackage{subfigure}
\usepackage{color}
\usepackage{graphicx}
\usepackage{braket}
\usepackage{bm}

\numberwithin{equation}{section}

\newcommand{\be}{\begin{equation}}
\newcommand{\ee}{\end{equation}}
\newcommand{\nn}{\nonumber}
\newcommand{\f}{\frac}
\newcommand{\p}{\partial}

\newcommand{\tj}[6]{ \begin{pmatrix}
       #1 & #2 & #3 \\
       #4 & #5 & #6 
\end{pmatrix}}

\newtheorem*{proposition*}{Proposition}

\theoremstyle{remark}

\DeclareMathOperator{\im}{\mathrm{i}}

\let\a=\alpha \let\b=\beta  \let\g=\gamma  \let\d=\delta
\let\z=\zeta        \let\l=\lambda
\let\m=\mu              \let\r=\rho

\let\G=\Gamma \let\D=\Delta    \let\X=F
         
  \let\eps=\epsilon

\newcommand{\gb}{\bar{g}}

\newcommand{\cA}{\mathcal{A}}

\newcommand{\cF}{\mathcal{F}}
\newcommand{\cG}{\mathcal{G}}

\newcommand{\cO}{\mathcal{O}}

\newcommand{\cZ}{\mathcal{Z}}

\title{\bf Remarks on a melonic field theory\\ with cubic interaction
}

\author[1]{Dario Benedetti}
\author[2]{and Nicolas Delporte}

\affiliation[1]{CPHT, CNRS, Ecole Polytechnique, Institut Polytechnique de Paris, Route de Saclay, \\ 91128 Palaiseau, France}
\affiliation[2]{Okinawa Institute of Science and Technology Graduate University, \\ 1919-1 Tancha, Onna, Kunigami, Okinawa, Japan 904-0412}

\emailAdd{dario.benedetti@polytechnique.edu}
\emailAdd{nicolas.delporte@oist.jp}

\abstract{
We revisit the Amit-Roginsky (AR) model in the light of recent studies on Sachdev-Ye-Kitaev (SYK) and tensor models, with which it shares some important features. It is a model of $N$ scalar fields transforming in an $N$-dimensional irreducible representation of $SO(3)$. The most relevant (in renormalization group sense) invariant interaction is cubic in the fields and mediated by a Wigner $3jm$ symbol. The latter can be viewed as a particular rank-3 tensor coupling, thus highlighting the similarity to the SYK model, in which the tensor coupling is however random and of even rank. As in the SYK and tensor models, in the large-$N$ limit the perturbative expansion is dominated by melonic diagrams. The lack of randomness, and the rapidly growing number of invariants that can be built with $n$ fields, makes the AR model somewhat closer to tensor models.
We review the results from the old work of Amit and Roginsky with the hindsight of recent developments, correcting and completing some of their statements, in particular concerning the spectrum of the operator product expansion of two fundamental fields. For $5.74<d<6$ the fixed-point theory defines a real CFT, while for smaller $d$ complex dimensions appear, after a merging of the lowest dimension with its shadow.
We also introduce and study a long-range version of the model, for which the cubic interaction is exactly marginal at large $N$, and we find a real and unitary CFT for any $d<6$, both for real and imaginary coupling constant, up to some critical  coupling.}

\begin{document}
\maketitle
\flushbottom

\section{Introduction}

Following the introduction of the SYK model \cite{Sachdev:1992fk,Kitaev,Polchinski:2016xgd,Maldacena:2016hyu}, and of tensor models with similar features \cite{Witten:2016iux,Klebanov:2016xxf}, there has been some interest in quantum field theories that  in the large-$N$ limit are dominated by melonic diagrams.
In this respect, tensor models have played a driving role, as they are genuine quantum field theories, whereas the SYK model is a disordered model with a random coupling. 
It is of course still possible to study SYK-like models in more than one dimension \cite{Turiaci:2017zwd,Berkooz:2017efq,Murugan:2017eto}, but tensor models have some more appealing features, in particular in view of a potential AdS/CFT correspondence, such as the fact that they have a global symmetry from the start, which could then be gauged, whereas in SYK-like models the symmetry only emerges after quenching.
As a consequence several tensor models have been studied, with a focus on the fact that their melonic large-$N$ limit allows the identification of non-trivial fixed points of the renormalization group and the non-perturbative computation of the spectrum of bilinear operators \cite{Klebanov:2016xxf,Giombi:2017dtl,Bulycheva:2017ilt,Prakash:2017hwq,Benedetti:2017fmp,Gubser:2018yec,Giombi:2018qgp,Benedetti:2018ghn,Popov:2019nja,Benedetti:2019eyl,Benedetti:2019ikb,Benedetti:2019rja,Benedetti:2020yvb} (see also \cite{Delporte:2018iyf,Klebanov:2018fzb,Gurau:2019qag,Benedetti:2020seh} for reviews and more references).

In this note, we wish to revisit an old model by Amit and Roginsky \cite{Amit:1979ev} which has a melonic large-$N$ limit, but which so far seems to have gone largely unnoticed in the high-energy community.\footnote{The model has attracted instead some attention in the context of nonlinear stochastic equations and of spin-glasses \cite{Mou_1993,Doherty_1994,Franz_1995}.} 
We will analyze its features with the hindsight of recent developments, correct some small mistakes in the original analysis, and provide some further results and generalizations of the model.

The Amit-Roginsky (AR) model has a number of interesting characteristics.
First of all, like tensor models, it is a genuine quantum field theory with a continuous global symmetry. The model is indeed invariant under field transformations in an $N$-dimensional irreducible representation of $SO(3)$.
Interactions are then expressed as products of $q$ fields $\phi_m$, with $m=1\ldots N$, contracted with an $SO(3)$-invariant tensors of rank $q$.
Interestingly, such symmetry allows a unique cubic invariant ($q=3$), the invariant tensor being given by the Wigner $3jm$ symbol, with $N=2j+1$. The cubic interaction is the most relevant in the renormalization group sense, and the main observation by Amit and Roginsky was that such interaction leads to a melonic large-$N$ limit.\footnote{They did not use the term ``melonic", which to the best of our knowledge was introduced in \cite{Bonzom:2011zz}.}
Therefore, the AR model provides a so far unique case of quantum field theory with melonic limit having a cubic interaction, as tensor models admitting a melonic limit have always interactions with an even number of fields.

Theories with cubic interactions have been studied since the early days of the renormalization group: the beta functions for a generic multiscalar model with cubic interactions have been computed at one loop in \cite{Ma:1975vn}, and for the case with a global symmetry such that there is a single coupling they have been computed at two loops in \cite{Amit:1976pz,McKane:1976zz,Mckane:1977bv}, at three loops in \cite{deAlcantaraBonfim:1980pe}, and at four loops in \cite{Gracey:2015tta}.
One important motivation, which was also the main one of Amit and Roginsky, comes from the Potts model, which in its field theory formulation has a cubic interaction \cite{Zia:1975ha,Amit:1976pz} (see also \cite{Zinati:2017hdy} and references therein), but much work has gone into models with cubic interactions for many other reasons, e.g.\ \cite{Cardy:1976ps,deAlcantaraBonfim:1981sy,collins_1984,Bellon:2009ju,Fei:2014xta,Fei:2014yja,Osborn:2017ucf,Codello:2019isr,Gracey:2020baa,Gracey:2020tkk,Bellon:2020qlx}.
It is therefore interesting that a melonic limit can be realized in a theory with cubic interaction.

We also observe that in the light of the results of \cite{Benedetti:2019sop}, we could view the Amit-Roginsky model as an on-shell version (or saddle-point approximation) of a bosonic SYK-like model with quenched disorder, i.e.\ with a randomly distributed rank-3 tensor coupling in place of the $3jm$ symbol. The distribution would need to be non-Gaussian, and with at least a ``pillow" or ``tetrahedron" quartic term and a negative coupling for the quadratic term, in order to allow a non-trivial solution, but that does not lead to crucial differences with respect to the Gaussian case, as shown in Ref.~\cite{Krajewski:2018lom}.
On the other hand, there is  one other aspect for which the AR model is closer to tensor models than to the SYK model, besides it having no random coupling: being invariant under an $N$-dimensional irreducible representation of $SO(3)$, rather than the fundamental of $O(N)$, it has many more invariants than just the simple bilinears of a vector model. And they grow very rapidly with the number of fields $q$: as we said, for $q=3$ there is only one invariant, while for $q=4$ there are already $N$ invariants.

Lastly, we notice that, like other models with a melonic limit, also the interaction of the AR model, being cubic in the fields, is unbounded from below.
This seems to be a universal feature of melonic theories, and it might explain the appearance of complex scaling dimensions in integer spacetime dimensions (for short-range models).
On the other hand, a priori the unboundedness is not necessarily a problem at large-$N$, or under other circumstances.
It has been argued (see for example \cite{deAlcantaraBonfim:1980pe} and references therein) that in the case of a model with just a cubic interaction the instability and its related problems can be avoided by taking an imaginary coupling \cite{Kirkham:1978wh}, as in the Lee-Yang model \cite{Fisher:1978pf,Cardy:1985yy}, or by taking  special limits, such as the $n\to 0$ limit of the $(n+1)$-state Potts model \cite{Houghton:1978dt}, as in the percolation problem \cite{Fortuin:1971dw}.
The large-$N$ limit can have a similar effect, at least near the upper critical dimension, as we will see below. In this respect, explicit calculations in the cubic $O(N)$ model of \cite{Fei:2014xta,Fei:2014yja} have shown that indeed imaginary parts of scaling dimensions are (exponentially) suppressed at large $N$ \cite{Giombi:2019upv}.
Moreover, along the lines of long-range $O(N)^3$ tensor model \cite{Benedetti:2019eyl,Benedetti:2019ikb,Benedetti:2019rja,Benedetti:2020yvb}, we will introduce and study also a long-range version of the AR model, for which the coupling is exactly marginal at large $N$, and we will find that in this case a real and unitary CFT can be identified at small coupling for either real or imaginary coupling, even at integer dimensions $d<6$.

\paragraph{Plan of the paper.}
We begin in Section \ref{sec:model} with the definition of the short-range and long-range versions of the AR model, and a discussion of its Feynman diagrams and large-$N$ limit.
In Section \ref{sec:SD} we study the melonic Schwinger-Dyson equations for the two-point function, and in particular we recognize the generating function of 3-Catalan numbers in the coefficient of the solution in the long-range case.
In Section \ref{sec:beta-largeN} we confirm by standard RG methods the existence of the large-$N$ fixed point, while in Section \ref{sec:finiteN} we consider the finite-$N$ corrections.
Lastly, in Section \ref{sec:spectrum} we study the spectrum of bilinear operators, i.e.\ the operators appearing in the operator product expansion of two fundamental fields. As to that end we use the conformal partial wave expansion of the four-point function and we take the chance to discuss in Appendix \ref{app:CPW} a small subtlety that arises in the identification of the physical spectrum for melonic theories with higher-order interactions. We summarize our findings in Section \ref{sec:concl}.

\section{The Amit-Roginsky model and its long-range version}
\label{sec:model}

The Amit-Roginsky (AR) model, introduced in Ref.~\cite{Amit:1979ev}, is a bosonic model of $N$ complex scalar fields $\phi_m$, with $m=1\ldots N$, in an irreducible representation of $SO(3)$ of dimension $N=2j+1$, and with a cubic interaction mediated by a Wigner $3jm$ symbol. 
We will consider a slight variation of the model, choosing real scalars and allowing the quadratic part of the action to be long range.
The action reads:
\be \label{eq:AR}
\begin{split}
S = \int d^d x \; &\left(\f12 \sum_m \left(\phi^m (-\p^2)^\z\phi_m + \l_2\, \phi_m \phi^m\right) \right.\\
&\left.+\sum_{m_1,m_2,m_3} \f{\l}{3!}\sqrt{2j+1}\tj{j}{j}{j}{m_1}{m_2}{m_3} \phi^{m_1}\phi^{m_2}\phi^{m_3} \right) \,,
\end{split}
\ee
where $\tj{j}{j}{j}{m_1}{m_2}{m_3}$ is the $3jm$ symbol. As at equal $j$'s the latter vanishes for half-integer $j$, the model is restricted to integer $j$ (that is why we have a representation of $SO(3)$ rather than of $SU(2)$), i.e.\ odd $N$.
Moreover, since the $3jm$ symbol is antisymmetric for odd $j$ and symmetric for even $j$, we must restrict to even $j$ for a non-vanishing interaction.
For the representation theory of $SO(3)$ we mostly follow the notation and conventions of Ref.~\cite{Yutsis:1962vcy}; see also \cite{Benedetti:2019sop} for a brief list of useful formulas and conventions.
Indices are raised and lowered by the invariant metric, defined as:
\be \label{eq:metric_j}
g_{j}^{m m'} = g^{j}_{m m'} \equiv \sqrt{2j+1} \tj{j}{j}{0}{m}{m'}{0} = (-1)^{j-m} \d_{m\, -m'} \,.
\ee
Notice that the invariant metric is its own inverse, i.e.\ $\sum_{m''} g^{j}_{m m''} g_{j}^{m'' m'} = \d_m^{m'}$, and that the $3jm$ symbol is traceless, i.e.:
\be \label{eq:3jm-trace}
\sum_{m_2, m_3} \tj{j}{j}{j}{m_1}{m_2}{m_3} g_j^{m_2 m_3} = 0\,.
\ee
As a consequence of the latter identity, tadpole diagrams vanish identically.

The idea behind Amit and Roginsky's work was to generalize the Potts model (also described by a multiscalar theory with cubic interaction \cite{Zia:1975ha,Amit:1976pz}) by endowing it with a continuous symmetry, in such a way to allow the introduction of a useful large-$N$ limit. In this optic, a single complex scalar field (i.e.\ two real scalars) with a cubic interaction corresponds to the 3-state Potts model, with a discrete symmetry group (the dihedral group $D_3$); the continuous $SO(3)$ symmetry is then superimposed to it in order define the large-$N$ limit.
Here we are not interested in the connection to the Potts model, and thus we are free to choose real fields. We will briefly compare the real and complex versions in Section~\ref{sec:variants}.

Quartic interactions could be added to the model to stabilize the potential.
There are actually many of them, as there are $2j+1$ quartic invariants, of the form
\be
g_J^{m m'}\tj{j}{j}{J}{m_1}{m_2}{m} \tj{j}{j}{J}{m_3}{m_4}{m'} \phi^{m_1}\phi^{m_2}\phi^{m_3}\phi^{m_4}\,,
\ee
for any $J=0,1,\ldots,2j$.
However, for $4<d<6$ such terms are irrelevant in the IR, hence we do not include them. Moreover, it is rather common to consider unbounded potential at large $N$, as the instability might be suppressed in the limit.

As a generalization of the original model, we here allow for a long-range propagator, therefore introducing in the kinetic term a Laplacian to a power $\z$, which we take to be $0<\zeta\leq 1$, in order to preserve the thermodynamic limit and reflection positivity of the propagator. More concretely, we take as free propagator\footnote{Our convention for the Fourier transform is the same as in \cite{Giombi:2017dtl}, i.e. $\int dx e^{-ikx}f(x)=\hat{f}(k)$ and for convenience, we remove the hats.}
\be
C(p) = \f{1}{p^{2\zeta}}\,, \;\;\;\; C(x,y) = \int \f{d^d p}{(2\pi)^d}\f{e^{ip(x-y)}}{p^{2\zeta} }= \f{\G\left(\Delta_{\phi}\right)}{2^{2\z}\pi^{d/2}\G(\z)} \f{1}{|x-y|^{2\D_{\phi}}}\,.
\ee
The original (short-range) model is recovered for $\z=1$. The canonical dimension of the field is
\be
\D^{c}_\phi = \f{d-2\z}{2} \,.
\ee
For $\z=1$ the upper critical dimension is $d=6$, while for $\z=d/6$, we find $\D^{c}_\phi = d/3$, hence in such case the cubic term becomes marginal for any $d$.
In order to make the interaction slightly relevant, or to regularize the critical theory, we introduce a small parameter $\eps$, either via dimensional continuation at fixed $\z$ (i.e.\ $d=6\z-\eps$), or via analytical continuation of $\z$ at fixed $d$ (i.e.\ $\z=(d+\eps)/6$). In the short-range case we will employ the first continuation, as usual, while in the long-range case we will opt for the second.

\begin{figure}[htbp]
\centering
\includegraphics[width=0.45\textwidth]{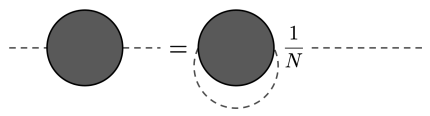}\\
\hspace{8mm}\includegraphics[width=0.45\textwidth]{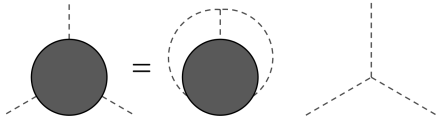}
\caption{Representation of the simplifying identities for the isospin structure of two- (top) and three-point (bottom) functions. Dashed edges correspond to contraction of $SO(3)$ indices via the invariant metric, the dark blobs represent the arbitrary internal structure of the diagrams, and the explicit 3-valent vertices in the reduction of the three-point function are associated to $3jm$ symbols.}
\label{fig:recoupling}
\end{figure}

\begin{figure}[htbp]
\begin{minipage}[c]{.6\textwidth}
\centering
\includegraphics[width=0.8\textwidth]{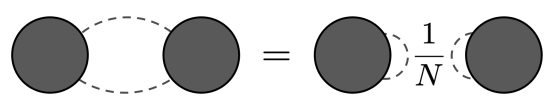}\\
\includegraphics[width=0.8\textwidth]{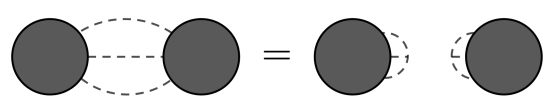}
\end{minipage}
\begin{minipage}[c]{.4\textwidth}
\centering
\includegraphics[width=0.4\textwidth]{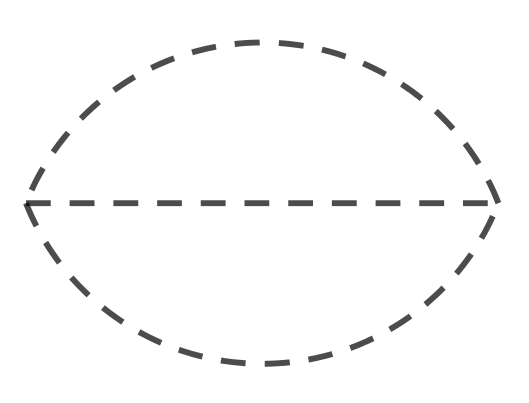}
\end{minipage}
\caption{Left: Factorization properties of 2PR (top) and 3PR (bottom) representations of isospin diagrams. Right: The melon diagram, i.e.\ the simplest 2-particle irreducible isospin diagram.}
\label{fig:factorization}
\end{figure}

\paragraph{Graphical technique and large-$N$ limit.}
Following standard Feynman rules, the propagator and vertex of the theory imply that the amplitude of any Feynman diagram $\g$ factors as
\be \label{eq:generic_amplitude}
\cA_\g = c_\g \left(\f{\l}{3!}\sqrt{N}\right)^{v(\g)}I_\g \, A_\g \,,
\ee
where $v(\g)$ is the number of vertices of $\g$, $c_\g$ is the combinatorial factor of the diagram, $I_\g$ the usual spacetime (or momentum) integral, and $A_{\g}$ is a purely group theoretic factor.
In order to understand the large-$N$ limit it is then useful to introduce two separate diagrammatic representations for keeping track of the two contributions $I_\g$ and $A_\g$ to a perturbative amplitude, which Amit and Roginsky called ``isoscalar'' (or spatial) and ``isospin'' contribution, respectively. We will use solid lines for the isoscalar diagrams and dashed lines for the isospin diagrams. In the latter, a (three-valent) vertex is associated to a $3jm$ symbol, and an edge to the invariant metric \eqref{eq:metric_j}.
A similar double representation is used also in tensor models, with the isospin diagrams replaced by edge-colored graphs (e.g.\ \cite{Benedetti:2020seh}).
However, while in tensor models the edge-colored graphs encode the internal structure of the invariants sitting at the vertices of the usual Feynman diagrams, and hence have a different topology from the latter, in the AR case the two types of diagrams have the same topology. Nevertheless, the isospin diagrams are useful in determining the factor of $N$ in a given amplitude because one can exploit the diagrammatic rules of $SU(2)$ recoupling theory \cite{Yutsis:1962vcy}, and thus perform contractions and other combinatorial operations that have no equivalent in the spatial part of the amplitude.
In particular, by using standard identities of recoupling theory, two- and three-point diagrams are proportional to the invariant metric and the $3jm$ symbol, respectively, as shown in Fig.\ref{fig:recoupling}, and we only need to be concerned with the proportionality factors, which are vacuum diagrams.
Moreover, two- and three-particle reducible (2PR and 3PR) isospin diagrams can be factorized as drawn in Fig.~\ref{fig:factorization}.

We call ``fully 2PR diagrams" those diagrams for which iterating the 2PR factorization leads to no other two-particle irreducible (2PI) diagrams than the simplest possible 2PI diagram, also known as the melon diagram, represented on the right of Fig.~\ref{fig:factorization}.
Since the latter has isospin amplitude $A_{\rm melon}=1$, it follows from Fig.~\ref{fig:factorization} that a fully 2PR vacuum diagram with $v=2n$ has isospin amplitude
\be
A_{\rm fully-2PR} = N^{1-n} \,.
\ee
Combined with the explicit factor of $N$ in \eqref{eq:generic_amplitude}, we conclude that fully 2PR diagrams are of order one.

By the factorization rules of Fig.~\ref{fig:factorization}, we can reduce the isospin amplitude of any other closed diagram with $2n$ vertices to the product
\be
A_{\g} = N^{-n_0} \prod_{i=1}^k A_{\{3n_i j\}} \,,\;\;\;\; n=1+n_0-k+\sum_{i=1}^k n_i\,,
\ee
where $\{3nj\}$ stands for a three-particle irreducible (3PI) diagram with $2n$ vertices, also known as $3nj$ symbol.
The general asymptotic behavior of $3nj$ symbols is an open problem,\footnote{See for example Ref.~\cite{Haggard:2009kv,Costantino2011,Bonzom:2011cy,Dona:2017dvf} and references therein.} but in order to determine the leading order diagrams of the AR model, a rough bound (in particular ignoring oscillating factors of order one) is sufficient.
By a combination of analytical evaluations (for $3nj$ symbols of first and second kind, i.e.\ those that can be written as a single sum of products of $6j$ symbols \cite{Yutsis:1962vcy}) and numerical estimates (for other $3nj$ symbols up to $n=6$), Amit and Roginsky concluded that $3nj$ symbols are always subleading with respect to fully 2PR diagrams:
\be \label{eq:AR-bound}
|\{3nj\}|\lesssim N^{-n+1-\a'}  \,,
\ee
for $N=(2j+1)\to\infty$, with $\a'>0$.
As three-valent  fully 2PR diagrams are equivalent to three-valent melonic diagrams \cite{Bonzom:2011zz}, we conclude that the large-$N$ limit of the AR model is dominated by melonic Feynman diagrams.

It should be stressed that we are currently still lacking a proper proof of the bound \eqref{eq:AR-bound}, valid for all $n$ and all kinds of $3nj$ symbols. However, the available results show its validity at least up to $n=6$, that is sufficient to prove the melonic dominance up to five loops ($3nj$ symbols appear for the first time at $n-1$ loops, see Section~\ref{sec:finiteN}).

\begin{figure}[htbp]
\centering
\includegraphics[width=0.25\textwidth]{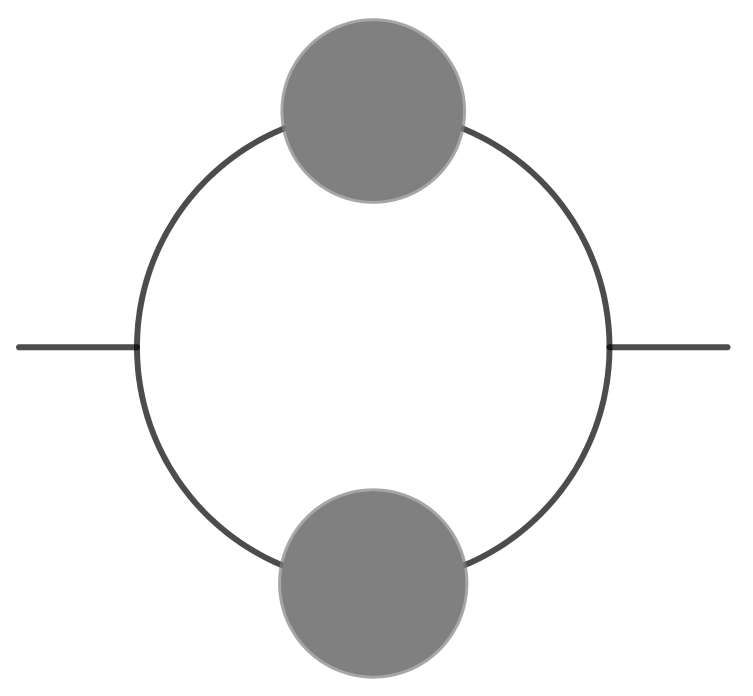}\\
\caption{The self-energy of the model. Full propagators are represented by the gray blobs.}
\label{fig:self-energy}
\end{figure}

\subsection{Other variants of the model}
\label{sec:variants}

One interesting variant of the model consists in taking complex fields, and writing the action
\be \label{eq:AR-compl}
\begin{split}
S_{1}= \int d^d x \; &\left(\sum_m \left(\bar\phi^m (-\p^2)^\z\phi_m + \l_2 \bar\phi_m \phi^m\right) \right.\\
&\left.+\sum_{m_1,m_2,m_3} \f{\l}{3!}\sqrt{2j+1}\tj{j}{j}{j}{m_1}{m_2}{m_3}( \phi^{m_1}\phi^{m_2}\phi^{m_3} + \bar\phi_{m_1}\bar\phi_{m_2}\bar\phi_{m_3} ) \right) \,,
\end{split}
\ee
which in fact, for $\z=1$, is the original AR model.
An interesting consequence of the complex nature of the fields is that Wick contractions are only possible between a $\phi$ and a $\bar\phi$, hence any Feynman diagram needs to be bipartite. That is, for any Feynman diagram of the theory it must be possible to separate its set of vertices $V$ into two subsets, $V_+$ and $V_-$, such that any vertex in one subset has only adjacent vertices from the other subset. 
In other words, diagrams containing a cycle with an odd number of edges are not allowed, and thus we have a reduced number of diagrams than in the real version of the model.\footnote{Such a reduction was first noticed in \cite{Mckane:1977bv} for a model with $SU(3)\times SU(3)$ symmetry.}
As at large-$N$ the theory is still dominated by melonic diagrams, which are bipartite, the difference between real and complex versions of the model only becomes manifest at subleading orders in $1/N$, as we will discuss in Section~\ref{sec:finiteN}.

Another variant can be considered along the lines of the bi-adjoint model of \cite{Gracey:2020baa,Gracey:2020tkk}, writing a two-index field $\phi_{m n}$
\be \label{eq:AR-matrix}
\begin{split}
S_{2}= \int d^d x \; &\left(\f12 \sum_{m,n} \left(\phi^{mn} (-\p^2)^\z\phi_{mn} + \l_2\, \phi_{mn} \phi^{mn}\right) \right.\\
&\left.+\sum_{\{m_i,n_i\}} \f{\l}{3!}(2j+1)\tj{j}{j}{j}{m_1}{m_2}{m_3}\tj{j}{j}{j}{n_1}{n_2}{n_3} \phi^{m_1 n_1}\phi^{m_2 n_2}\phi^{m_3 n_3} \right) \,.
\end{split}
\ee
The effect is to trivially double all the isospin diagrams, so it does not alter much the structure of the theory. However, odd $j$ is allowed in this case.

It would also be possible to let the two indices of $\phi_{m n}$ be in different representations, say $j_1$ and $j_2$, with both spin labels being even, or both odd, and both large. This could be interpreted as an $SO(4)$ invariant model, since $SO(4)$ is isomorphic to $SU(2)\times SU(2)$.

\section{Schwinger-Dyson equation for the two-point function}
\label{sec:SD}

In the large-$N$ limit, the melonic dominance leads to the standard melonic form for the Schwinger-Dyson equation:
\be \label{eq:SDeq}
G(p)^{-1} = Z  \left(p^{2\z} + \l_2 \right) - \f{\l^2}{2} \int_q G(q) G(p+q) \,,
\ee
where $\Sigma(p^2)=\f{\l^2}{2} \int_q G(q) G(p+q)$ is the melonic self-energy (cf. Fig.~\ref{fig:self-energy}).
We have included a field (or wave function) renormalization $Z$, in order to cancel divergences proportional to $p^{2\z}$ from the self-energy. However, notice the latter only occur for $\z=1$, as they are obtained by Taylor expanding $\Sigma(p^2)$ around $p=0$, and thus only include integer powers of $p^2$; in other words, this is the well-known statement that counterterms are local, and thus the non-local kinetic term of long-range models does not need renormalization.
The mass coupling should contain a counterterm canceling the $p$-independent divergent part of the self-energy, namely $Z \l_2 = \l_2^c + g_2$, with $\l_2^c = \f{\l^2}{2} \int_q G(q)^2$, which is zero in dimensional/analytic regularization. In the following we will ignore all such divergences which are zero in dimensional regularization.
Tuning the renormalized mass to the critical value $g_2 = 0$, one finds the solution $G(p)\sim p^{-d/3}$, valid in the IR limit, if $\z=1$, or at all scales, if $\z=d/6$.
The scaling form of the solution suggests the existence of an IR fixed point in the first case, and a line of fixed points in the second. In order to support such picture, one needs to consider the renormalization group flow of the coupling $\l$, which we do in the following section.

\paragraph{Short-range model ($\z=1$).}
In the case $\z=1$, we can solve the SD equation \eqref{eq:SDeq} approximately in the IR limit. The IR approximation amounts to discarding the $p^2$ term, and assuming the scaling ansatz $G(p)=  p^{-d/3} \cZ_{\rm SR}$, leading to
\be
\cZ_{\rm SR} = \left( \f{d (4\pi)^{d/2}}{3 \l^2} \frac{\G(d/6)^2\G(2d/3)}{\Gamma(d/3 )^2\Gamma( 1 - d/6)}\right)^{\f13} \,,
\ee
where we have used the identity 
\begin{equation} \label{eq:melon-int}
\int \frac{d^dq}{(2\pi)^d}\frac{1}{q^{2\a}(q+p)^{2\b}} = \frac{1}{(4\pi)^{d/2}}\frac{\Gamma(d/2 - \a)\Gamma(d/2 - \b)\Gamma(\a + \b - d/2)}{\G(\a)\G(\b)\G(d-\a-\b)}\frac{1}{p^{2(\a+\b - d/2)}}\,.
\end{equation}

\

\paragraph{Long-range model ($\z=d/6$).}
In the case $\z=d/6$, we can solve the SD equation \eqref{eq:SDeq} exactly. Setting $Z=1$ and $\l_2=0$ (or $\l_2=\l_2^c$ in a cutoff regularization), and assuming $G(p)= p^{-d/3} \cZ_{\rm LR}$, the SD equation reduces to an algebraic equation for $\cZ$:
\be \label{eq:calZ-eq}
1 = \cZ_{\rm LR} + \f{3 \l^2 }{d (4\pi)^{d/2}} \frac{\Gamma(d/3 )^2\Gamma(1 - d/6)}{\G(d/6)^2\G(2d/3)}\cZ^3_{\rm LR} \,.
\ee
The cubic equation can be solved exactly as a function of $\l$:
\be
\cZ_{\rm LR} = \frac{1}{6^{2/3}}\left(\sqrt{3} \sqrt{a^3 (27 a+4)}+9 a^2\right)^{\f13} \left(\frac{2^{1/3}}{a}-\frac{2 \times 3^{1/3}}{\left(\sqrt{3} \sqrt{a^3 (27 a+4)}+9
   a^2\right)^{2/3}}\right) \,,
\ee
where
\be
a =  \f{3 \l^2 }{d (4\pi)^{d/2}} \frac{\Gamma(d/3 )^2\Gamma( 1 - d/6)}{\G(d/6)^2\G(2d/3)} \,.
\ee
Notice that $a>0$ for real $\l$ and $d<6$, and that $\cZ_{\rm LR}$ is real for all positive values of $a$, and $\cZ_{\rm LR}\sim a^{-1/3}$ for $a\to+\infty$.
For imaginary $\lambda$ we have instead $a<0$, and a singular point is found at $a=-4/27$, where $\cZ_{\rm LR}$ reaches a finite value ($\cZ_{\rm LR}=3/2$), but with infinite slope.

Interestingly, $\cZ_{\rm LR}$ is the generating function of 3-Catalan (or Fuss-Catalan) numbers, that is its series expansion is
\be \label{eq:Catalan_series}
\cZ_{\rm LR} = \sum_n \f{1}{3n+1} \binom{3n+1}{n} (-a)^n \,.
\ee
The appearance of 3-Catalan numbers can be easily understood. In fact, with $C(p)=p^{-d/3}$, and by use of \eqref{eq:melon-int}, any melonic insertion (i.e.\ insertion of the one-loop two-point diagram in an edge) in a diagram $\g$ has the simple effect of multiplying the amplitude of the original diagram, $\cA_\g$, by a factor $-a$. Therefore, summing all the melonic two-point diagrams reduces to a known combinatorial problem, whose solution is captured by the generating function of Fuss-Catalan numbers \cite{Bonzom:2011zz}. 

It is also instructive and useful to solve equation \eqref{eq:calZ-eq} in terms of the rescaled\footnote{As here $\cZ_{\rm LR}$ is a finite quantity, we prefer to call this a rescaling rather than a renormalization.} coupling $g=\l \cZ^{3/2}_{\rm LR}$. The rescaling is an effective way of resumming the melonic two-point functions and absorbing their contribution in the coupling: once we switch to the coupling $g$, we only have to consider skeleton diagrams with no melonic insertions, and with $C(p)=p^{-d/3}$ as propagator.
In terms of the rescaled coupling, the solution reads
\be
\cZ_{\rm LR} = 1 - g^2/g_{c,+}^2 \,,
\ee
with
\be \label{eq:g_+}
g_{c,+}^2 = d (4\pi)^{d/2} \frac{\G(d/6)^2\G(2d/3)}{3 \Gamma(d/3 )^2\Gamma(1 - d/6)} 
=\f{\l^2}{a} = \f{\l^2}{\cZ_{\rm SR}^3} \,.
\ee
We notice that $g=g_{c,+}$ means $\cZ_{\rm LR}=0$, but also $\cZ_{\rm LR}^3=1/a$, or $\cZ_{\rm LR}=\cZ_{\rm SR}$, which is the solution obtained when we discard the inverse free propagator in the SD equation. The two equations for $\cZ_{\rm LR}$ are of course only consistent in the limit $\l\to\infty$.
On the other hand, for imaginary $\lambda$, and thus imaginary $g$, $\cZ_{\rm LR}$ stays finite and positive, but only up to the critical value 
\be
g_{c,-}^2 = g_{c,+}^2 \left(a\cZ_{\rm LR}^{3}\right)_{\big|_{a=-4/27}}  = - \f{1}{2} g_{c,+}^2 \,,
\ee
i.e.\ the critical point of $\cZ_{\rm LR}$, at which the relation between $g$ and $\lambda$ ceases to be invertible.

A similar picture was found in the melonic limit of the long-range $O(N)^3$ model in \cite{Benedetti:2019eyl}, with the only difference that, the interaction being quartic, the equation for the analogue of our $\cZ_{\rm LR}$ was quartic in that case and its solution is the generating function of 4-Catalan numbers.


\section{Large-$N$ beta functions and fixed points}
\label{sec:beta-largeN}

\paragraph{Short-range model ($\z=1$).}

We define the renormalized dimensionless coupling $g$ via
\be
\l = \m^{\eps/2} Z^{-3/2} g \,,
\ee
with $Z$ the wave-function renormalization, which we can fix by the renormalization condition
\be
\lim_{\epsilon\to 0}\frac{d\Gamma^{(2)}(p)}{dp^2}|_{p^2=\mu^2}=1\,,
\ee
where $\Gamma^{(2)}(p)=G(p)^{-1}$.

Since at leading-order in the large-$N$ limit there is no vertex correction, the beta function reads
\be \label{eq:beta_largeN}
\b(g) = \f{g}{2}(-\eps + 3\eta(g) )\,,
\ee
where we defined the varying anomalous dimension $\eta(g) = \m\p_\m \ln Z$.
At a non-trivial fixed point, i.e.\ at $g=g^\star\neq 0$ such that $\b(g^\star)=0$, we necessarily have $\eta^\star\equiv \eta(g^\star) = \eps/3$.
Remembering that at a fixed point the field dimension is
\be
\D_\phi = \f{d-2+\eta^\star}{2} \,,
\ee
we see that for $d=6-\eps$ and $\eta = \eps/3$ we recover $\D_\phi= d/3$, i.e.\ $G(p)\sim p^{d/3}$ as expected from the SD equation.

The question is for what value of $g$, if any, does the anomalous dimension equal $\eps/3$.
Such question was addressed indirectly in \cite{Amit:1979ev}, by demanding that a slow transient in the IR solution of the SD equation cancel. We will do a more standard computation here.

Considering the SD equation at one loop, we have to perform again the melonic integral of Fig.~\ref{fig:self-energy}, this time using the bare propagator with $\z=1$; we find that $Z$ obeys the equation 
\be 
1 = Z - \frac{\l^2 \m^{-\eps}}{Z^22(4\pi)^3}\frac{\Gamma(\eps/2 - 1)}{3!} + \cO(\l^2\eps^0)\,,
\ee 
which leads to 
\be
Z = 1 - \frac{\l^2 \m^{-\eps}}{6(4\pi)^3 \eps} + \cO(\l^4)\,.
\ee
From this, we conclude that 
\be \label{eq:eta_largeN}
\eta = \frac{\mu^\eps \l^2}{6(4\pi)^3} + \cO(\l^4)  = \frac{ g^2}{6(4\pi)^3} + \cO(g^4) \,.
\ee

The fixed point condition $\eta=\epsilon/3$ gives two solutions:
\be \label{eq:FP-SR}
g^\star_\pm = \pm 8 \sqrt{2\pi^3 \epsilon} + \cO\left(\epsilon^{3/2}\right) \,,
\ee
in agreement with what was obtained in \cite{Amit:1979ev}. Notice that since the cubic interaction is unbounded from above and from below, we have no restriction on the sign of the coupling, and hence the two solutions have the same status.
The critical exponent describing the approach to the fixed point, also known as correction-to-scaling exponent, is given by
\be
\omega = \b'(g^\star_\pm) = \epsilon + \cO\left(\epsilon^{2}\right)\,,
\ee
which is positive, as expected for an IR fixed point.
The dimension of the corresponding cubic operator is $\D_{\phi^3}=d+\omega = 6+ \cO\left(\epsilon^{2}\right)$.

\paragraph{Long-range model ($\z<1$).}

In the long-range model, with $\z=(d+\eps)/6<1$, there is no wave function renormalization because the kinetic term is non-local, while UV divergences always lead to local counterterms.
More explicitly, we have seen in the previous subsection that at $\epsilon=0$ the full-two point function is found to be proportional to the bare one, with finite proportionality factor $\cZ_{\rm LR}$.
Therefore, the anomalous dimension vanishes, and in the large-$N$ limit we simply have
\be
\b(g)=-\eps g/2\,.
\ee
At $\eps=0$, i.e.\ for $\z=d/6$, the beta function vanishes identically, hence the interaction is exactly marginal.
Such model thus defines a one-parameter family of conformal field theories (or a one-dimensional conformal manifold), similarly to the melonic limit of long-range tensor models with quartic \cite{Benedetti:2019eyl} and sextic interactions \cite{Benedetti:2019rja}. However, in the case of tensor models, there are several quartic or sextic couplings and only one of them has vanishing beta function, hence one needs to look for fixed points of the other beta functions. In the AR model instead there is only one cubic interaction, hence there are no other beta functions to consider; the situation resembles in this sense that of the $O(N)$ model with $(\phi^2)^3$ interaction, which is the only sextic interaction, and which at $d=3$ and in the large-$N$ limit is exactly marginal.

\section{Finite-$N$ beta functions and fixed points}
\label{sec:finiteN}

The beta functions of the short-range AR model at finite $N$ can be obtained as a special case of those for general multiscalar models with only one cubic coupling \cite{Amit:1976pz,Mckane:1977bv,deAlcantaraBonfim:1980pe,Gracey:2015tta,Gracey:2020tkk}.
The starting point of such calculations is an action like in \eqref{eq:AR}, but $\sqrt{2j+1}\tj{j}{j}{j}{m_1}{m_2}{m_3}$ replaced by a tensor $d_{m_1 m_2 m_3}$, which is assumed to be the only rank-3 invariant of some underlying symmetry group.
The latter assumption implies that any three-point function must be proportional to  $d_{m_1 m_2 m_3}$, as is the case in the $SO(3)$ case of the AR model (see Fig.~\ref{fig:recoupling}).

At two loops, one only needs to introduce the following proportionality coefficients (scalar invariants of the symmetry group, or Casimirs):
\be
 \sum_{\{m'_i\}} d_{m_1 m'_1 m'_2} d^{m_2 m'_1 m'_2} = T_2\, \d_{m_1}^{m_2}\,,
\ee
\be
 \sum_{\{m'_i\}} d_{m_1 m'_1}{}^{m'_2} d_{m_2 m'_2 }{}^{m'_3} d_{m_3 m'_3}{}^{m'_1}  = T_3\, d_{m_1 m_2 m_3}\,,
\ee
\be
 \sum_{\{m'_i\}} d_{m_1 m'_1 m'_2} d_{m_2 m'_3 m'_4} d_{m_3 m'_5 m'_6} d^{m'_1 m'_3 m'_5} d^{m'_2 m'_4 m'_6}  = T_5\, d_{m_1 m_2 m_3}\,.
\ee

Substituting $d_{m_1 m_2 m_3}=\sqrt{2j+1}\tj{j}{j}{j}{m_1}{m_2}{m_3}$, we find
\be \label{eq:Casimirs}
T_2=1\,, \;\;\; T_3 = (2j+1) \{6j\}\,, \;\;\; T_5 = (2j+1)^2 \{9j\}\,,
\ee
where we used the notation $\{6j\}$ and $\{9j\}$ as a shorthand for the $6j$ and $9j$ symbols with equal $j$'s, and we have used standard formulas \cite{Varshalovich:1988ye}, for even $j$.
Similarly, at three loops one finds only two new Casimirs, corresponding to the two kinds of $12j$ symbols, and at four loops five new Casimirs, corresponding to the five kinds of $15j$ symbols \cite{Yutsis:1962vcy}.\footnote{Notice that in  \cite{Gracey:2015tta} Gracey lists nine Casimirs at four loops, but the fact that only five of them are independent is generic, relying only on the fact that there exist only five topologies of cubic three-particle-irreducible (3PI) vacuum diagrams with ten vertices (up to a factor $N$, the Casimirs are obtained by contracting the three-point structure of the diagram with a $d^{m_1m_2m_3}$). Three of them do not have a unique representation as decagons plus internal edges, but the different representations are related by permutation of the vertices that leave the decagon structure intact (see discussion on $15j$ symbols in \cite{Yutsis:1962vcy}). In his notation we find $T_{93}=T_{95}=T_{97}$ and $T_{92}=T_{96}=T_{98}$.}

Equations \eqref{eq:Casimirs} should be substituted in the beta functions from \cite{Amit:1976pz,Mckane:1977bv,deAlcantaraBonfim:1980pe,Gracey:2015tta,Gracey:2020tkk}, which at two loops read:
\be \label{eq:beta_finiteN}
\b(\gb) = -\f{\eps}{2} \gb +\left(\f{T_2}{4}-T_3\right) \gb^3  +\f{1}{144}\left(-11 T_2^2+66 T_2 T_3 -108 T_3^2 -72 T_5\right) \gb^5 +\cO(\gb^7) \,,
\ee
where the bar stands for the rescaled coupling $\gb^2 = g^2 S_d/(2\pi)^d$, with $S_d=2\pi^{d/2}/\G(d/2)$ the area of the $(d-1)$-sphere with unit radius.

At large $j$, up to oscillating $O(1)$ factors, the $3nj$ symbols appearing at this order behave asymptotically as $|\{6j\}|\approx (2j+1)^{-3/2}$ \cite{Wigner:1959,PonzanoRegge} and $|\{9j\}|\approx (2j+1)^{-2-\a}$ with $1/2<\a<1$ \cite{Amit:1979ev,Haggard:2009kv}.\footnote{For generic values of the nine spins $j$, each rescaled by a factor $k\to\infty$, one would have $|\{9j\}|\approx k^{-3}$ \cite{Varshalovich:1988ye}, but the case with equal $j$'s corresponds to a degenerate configuration (a ``caustic'') in which the general asymptotic formula does not apply and a slightly slower decay is found \cite{Haggard:2009kv}.}
Therefore, the $T_3$ and $T_5$ contributions are subleading at large $N$, and and a similar conclusion holds for the  $12j$ and $15j$ symbols appearing at three and four loops, as can be checked numerically.
We can thus use the finite-$N$ calculations to extend our large-$N$ beta functions to four loops, setting $T_2=1$ and all the other Casimirs to zero in the results of  \cite{Gracey:2015tta}.\footnote{The beta function of  \cite{Gracey:2015tta} needs to be corrected by mapping $g^2\to-g^2$ and multiplying the coefficients by a factor 2. We thank John Gracey for explanations on this point.}
We find:
\be
\b(\gb) = -\f{\eps}{2} \gb +\f{1}{4} \gb^3  - \f{11}{144}  \gb^5 + \f{821}{20736}  \gb^7 - \f{20547}{746496}  \gb^9 +\cO(\gb^{11}) \,,
\ee
which at one loop agrees with \eqref{eq:beta_largeN} and \eqref{eq:eta_largeN}, after the rescaling of the coupling.
The fixed point and critical exponent are
\be
\gb^\star_\pm = \pm \left(\sqrt{2\epsilon } +\frac{11\, \epsilon^{3/2}}{18 \sqrt{2}} +\frac{13 \, \epsilon^{5/2}}{648 \sqrt{2}}+ \frac{623 \epsilon^{7/2}}{7776 \sqrt{2}} \right) +\cO \left(\epsilon^{9/2}\right)\,,
\ee
and
\be
\omega =\beta'(\gb^\star_\pm)  = \epsilon -\frac{11\, \epsilon^2}{18}+\frac{337\, \epsilon^3}{648} -\frac{16013 \epsilon^4}{23328}+\cO \left(\epsilon^5\right) \,.
\ee
\

Going back to finite $N$, with even $j$, we find that, with the exception of $j=6$ (i.e.\ $N=13$), the sign of $T_2/4-T_3$ is always positive. Therefore, we always have a real IR fixed point of order $\sqrt{\epsilon}$:
\be
\gb^\star_\pm = \pm \left(\f{2 \epsilon}{T_2-4T_3}\right)^{1/2} + \cO(\epsilon)\,,
\ee
with exponent
\be
\begin{split}
\omega =\beta'(\gb^\star_\pm) & = \f{\epsilon}{2} \left( \f{3}{T_2-4T_3}-1\right) +\cO \left(\epsilon^2\right) \\
&\simeq \epsilon \left( 1 + 6 \f{2^{5/4}}{\sqrt{\pi N}} \cos\left(3N \arccos\left(-1/3\right)+\f{\pi}{4}\right) +\cO(N^{-3/2}) \right) +\cO \left(\epsilon^2\right) \,,
\end{split}
\ee
where we used the Ponzano-Regge formula for the asymptotic expansion of the $6j$ symbol \cite{PonzanoRegge}.
Therefore, the finite-$N$ corrections do not spoil the existence of the fixed point found in the preceding section, except at $N=13$.
This should be contrasted with what happens in melonic theories with quartic interactions, in which a one-loop term proportional to $g^2$ is suppressed at large-$N$ with respect to a two-loop term proportional to $g^3$, and one has to assume that $\sqrt{\epsilon}\gg 1/N$ \cite{Benedetti:2020sye}; no similar assumption is needed in the case of a cubic interaction.

For the long-range model instead the finite-$N$ corrections have a drastic effect, as the beta function is no longer identically zero. In order to find a finite-$N$ precursor of the line of fixed points found at large-$N$, one has to introduce $\eps>0$ and use a double scaling limit with $\eps\sqrt{N}\equiv \tilde{\eps}\ll 1$, analogously to what was done for the long-range $O(N)^3$ model with quartic interaction \cite{Benedetti:2020sye}, or for the short-range $O(N)$ model with sextic interaction \cite{Fleming:2020qqx}.

Going back to the short-range model, we observe that in the case of a purely imaginary coupling, $\gb=\im \hat{g}$, the beta function for $\hat{g}$ has the opposite sign for the cubic term in \eqref{eq:beta_finiteN}, and thus an IR fixed point with real $\hat{g}\sim\sqrt{\epsilon}$ is only found at $j=6$.

Lastly, a small remark about the version of the model with complex fields: in this case, diagrams with cycles of length three (triangles) are to be excluded, which effectively amounts to setting $T_3=0$. At higher loops, only one $12j$ and one $15j$ symbols survive.
From the point of view of the large-$N$ expansion, the consequence is that  $1/N$ corrections in the beta function only start at two loops, with the $T_5$ term.

\section{Spectrum of bilinear operators}
\label{sec:spectrum}

One interesting, and much exploited, feature of the melonic limit is the possibility of deriving the full spectrum of operators which appear in the operator product expansion (OPE) of two fundamental fields. Such operators typically are schematically of the form $\phi (\p^2)^n \p_{\m_1}\ldots\p_{\m_J} \phi$, i.e.\ they are bilinear in the fundamental fields with an arbitrary number of derivatives, the uncontracted ones endowing the operator with spin $J$.

Let us briefly recall the theoretical background for the derivation of such OPE spectrum (see also \cite{Dobrev:1975ru,Caron-Huot:2017vep,Simmons-Duffin:2017nub,Karateev:2018oml} for the general theory, and \cite{Liu:2018jhs,Gurau:2019qag,Benedetti:2019ikb,Benedetti:2020seh} for applications to melonic CFTs).
First, we define  the forward four-point function, i.e. the part of the  four-point function which is connected in the $s$-channel ($12\to34$):
\be \label{eq:4pt-fwd}
\begin{split}
\cF_{ m_1 , m_2   m_3 , m_4 }(x_1,x_2,x_3,x_4) 
= & \Braket{ \phi_{m_1}(x_1) \phi_{m_2}(x_2) \phi_{m_3}(x_3) \phi_{m_4}(x_4)} \\
&\; - \Braket{ \phi_{m_1}(x_1) \phi_{m_2}(x_2)} \Braket{ \phi_{m_3}(x_3) \phi_{m_4}(x_4)} \\
= & \Braket{ \phi_{m_1}(x_1) \phi_{m_2}(x_2) \phi_{m_3}(x_3) \phi_{m_4}(x_4)}_c  \\
&\; +  \Braket{ \phi_{m_1}(x_1)  \phi_{m_3}(x_3) }\Braket{\phi_{m_2}(x_2) \phi_{m_4}(x_4) }\\ 
&\; +\Braket{ \phi_{m_1}(x_1) \phi_{m_4}(x_4) }\Braket{\phi_{m_2}(x_2) \phi_{m_3}(x_3)} \,.
\end{split}
\ee
The forward four-point function can be written as a conformal partial wave expansion \cite{Dobrev:1975ru,Simmons-Duffin:2017nub}:\footnote{Since we are considering a theory with cubic interaction we should in principle consider the part of $\cF$ which is one-particle irreducible in the $s$-channel, as done in \cite{Dobrev:1975ru}. However, due to the choice of index contractions we are looking at in \eqref{eq:4pt}, and to the traceless property \eqref{eq:3jm-trace} of the $3jm$ symbol, one-particle reducible in the $s$-channel do not contribute to it.}
\begin{equation} \label{eq:4pt}
\begin{split}
\sum_{m,m'} \cF_{m}{}^{ m}{}_{; m'}{}^{ m'}(x_1,x_2,x_3,x_4) 
 = N &\sum_{J\geq 0} 
  \int_{\frac{d}{2}-\im \infty}^{\frac{d}{2}+\im\infty} \frac{dh}{2\pi \im} 
  \;\frac{1}{1-k(h,J)} \; \mu^{\Delta_{\phi}}_{h,J}
     \cG^{\Delta_{\phi}}_{h,J}(x_i)\\
     & + (\text{non-norm.})\,,
\end{split}
\end{equation}
with $\cG_{h,J}(x_i) $ the conformal block, $\mu_{h,J}$ the measure, and $k(h,J)$ the eigenvalues of the two-particle irreducible four-point kernel, or Bethe-Salpeter kernel \cite{Dobrev:1975ru}.
The latter can for example be constructed from the 2PI effective action \cite{Benedetti:2018goh}.
The non-normalizable contributions are due to operators with dimension $h<d/2$, and they should be treated separately \cite{Simmons-Duffin:2017nub}. We will discuss them in more detail in App.~\ref{app:CPW}.

Closing the contour to the right, we pick poles at $k(h,J)=1$ (other poles coming from the measure and the conformal block are spurious and they cancel out \cite{Simmons-Duffin:2017nub}), and we recover an OPE in the $s$-channel:
\begin{equation}  \label{eq:4pt-OPE}
  \sum_{m,m'} \cF_{m}{}^{ m}{}_{; m'}{}^{ m'}(x_1,x_2,x_3,x_4) 
 = N \sum_{n,J} c_{n,J}^2  \; \cG^{\Delta_{\phi}}_{h_{n,J},J}(x_i) \,,
\end{equation}
where the dimensions of spin-$J$ operators, $h_{n,J}$, are the poles of $(1-k(h,J))^{-1}$, and the squares of the OPE coefficients $c_{n,J}$ are the residues at the poles.
Therefore, studying the four-point kernel we can obtain the spectrum of operators that appear in the OPE of two fundamental fields.

\begin{figure}[htbp]
\centering
\includegraphics[width=.6\textwidth]{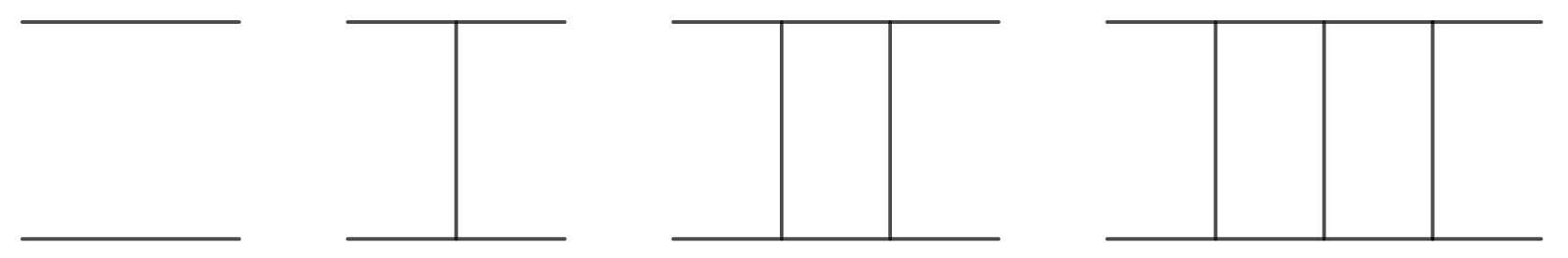}\\
\caption{The first four contributions to the ladder expansion of four-point function in the $s$-channel.}
\label{fig:1overNladder}
\end{figure}

The decomposition \eqref{eq:4pt} is generic and can be derived by means of the 2PI formalism; however, the explicit expression of $k(h,J)$ is typically only known in the large-$N$ limit.
In melonic theories with a $q$-valent interaction, the four-point function is a sum of ladder skeleton diagrams with rungs made of $q-2$ edges connecting the same two vertices. In the case of our cubic interaction, we get the simple ladders of Fig. \ref{fig:1overNladder}. The four-point kernel corresponds to the right-amputated single-rung ladder, and the sum over ladders is obtained as a geometric series of kernel convolutions.
The simple structure of the kernel in such case allows to extract the conformal dimensions of the bilinear operators as solutions of the equation $k(h,J)=1$.

\paragraph{Short-range model ($\z=1$).}

With the help of a technique that is by now standard, and in agreement with Ref. \cite{Amit:1979ev}, the spin-zero eigenvalues of the ladder kernel take the form 
\be  \label{eq:k-SR}
k(h,0) = -2 \frac{\G(d/6)\G(2d/3)\G(d/3-h/2)\G(h/2-d/6)}{\G(-d/6)\G(d/3)\G(2d/3-h/2)\G(h/2+d/6)} \,.
\ee 
We made use of the fact that, by conformal invariance, the four-point kernel $K$ satisfies the general eigenvalue equation
\be
k(h,0) v_h(x_1,x_2,x_3)= \int d^d y d^d z v_h (x_1,y,z)K(x_2,x_3,y,z) \,,
\ee
where $v_h(x_1,x_2,x_3)$ is a function with the conformal structure of a three-point function in a $d$-dimensional conformal field theory, between operators $O_h$ and $\phi^m$ of conformal dimension $h$ and $d/3$, respectively:
\be
v_h(x_1,x_2,x_3):=\expval{O_h(x_1)\phi^m(x_2)\phi_m(x_3)} = \f{C_{O_h\phi\phi}}{(x^2_{12}x^2_{13})^{h/2}(x^2_{23})^{d/3-h/2}} \,,
\ee 
where we denoted as usual $x_{ij} = x_i -x_j$. 
In order to extract \eqref{eq:k-SR} we have then used the triangle-vertex integral
\be
\int d^dx_0 \f{1}{x_{01}^{2\a_1}x_{02}^{2\a_2}x_{03}^{2\a_3}} =\f{\pi^{d/2}\Gamma(d/2-\a_1)\Gamma(d/2-\a_2)\Gamma(d/2-\a_3)}{\G(\a_1)\G(\a_2)\G(\a_3)(x^2_{12})^{d/2-\a_3}(x^2_{23})^{d/2-\a_1}(x^2_{13})^{d/2-\a_2}}\,,
\ee 
valid when $\a_1+\a_2+\a_3=d$.
The expression \eqref{eq:k-SR} is the special case with $q=3$ ot the general expression obtained in \cite{Giombi:2017dtl} for general melonic theories with $q$-valent interactions.

As explained above, we want to solve the equation $k(h,0)=1$. 
Setting $d=6-\eps$, we rewrite $h_n = 2\D_\phi^{(0)} + 2n + z$, where $\D_\phi^{(0)} = 2$ is the classical dimension of $\phi$ at $\epsilon=0$, and in order to find analytical solutions we expand $z$ in $\epsilon$, and subsequently we expand also $k(h_n,0)$, to then solve the equation order by order.
We find an infinite set of solutions
\begin{align}
h_{-1}&=2 + \f{5}{3}\eps + \f{19}{18} \eps^2 = \D_\phi + \eps + \f{19}{18} \eps^2 +\cO(\eps^3)\,,\\
h_0&=4 - \f{8}{3}\eps -\f{19}{18}\eps^2 = 2 \D_\phi -2\eps  -\f{19}{18}\eps^2  +\cO(\eps^3)\,,\\
h_1&=6 -\f{11}{18}\eps^2 = 2 \D_\phi +2 + \f{2}{3}\eps -\f{11}{18}\eps^2  +\cO(\eps^3)\,,\\
h_n&=4 + 2n - \f{2}{3}\eps + \f{4(n-2)!}{3(n+2)!}\eps^2 =  2 \D_\phi +2n + \f{4(n-2)!}{3(n+2)!}\eps^2 +\cO(\eps^3) \qquad (n\geq 2)\,.
\end{align}
We  expressed the operator dimensions also in terms of the field dimension at $\eps>0$, $\D_\phi=2-\eps/3$, to highlight their anomalous dimension, as $2\D_\phi+2n$ is the classical dimension of operators of the form $\phi\p^{2n} \phi$.
Since $h_{-1}<d/2$, the first solution is not met when moving the contour in \eqref{eq:4pt} to the right, hence it should not be included in the spectrum: it corresponds to the shadow operator \cite{Ferrara:1972uq} of the $\phi^2$ operator, with dimension $h_{-1}=d-h_0$; we  elaborate further on this in App.~\ref{app:CPW}.\footnote{One could be mislead to interpret $h_{-1}$ as the dimension of $\phi$, expecting it to appear in the OPE of two fundamental fields because of the cubic interaction. However, this is incorrect for two reasons: first, the dimension of $\phi$ at the fixed point is constrained to be $\D_\phi=2-\eps/3\neq h_{-1}$; second, the three-point function $\Braket{\phi_m \phi^m \phi_{m'}}$ is zero due to the traceless property \eqref{eq:3jm-trace} of the $3jm$ symbol.}
We remark that the solutions with $n\geq 1$ were missed in \cite{Amit:1979ev}, and moreover $h_{-1}$ was mistakenly taken to be the only physical solution. 

Having a full expression for $k(h,0)$, one can go beyond the $\eps$ expansion, and compute the spectrum numerically in arbitrary dimension. However, it turns out that very soon the spectrum becomes complex.
By a numerical solution of the $k(h,0)=1$ equation, we find that at $\eps \approx 0.264$, $h_0$ merges with $h_{-1}$ at $h_0=d/2$, and then they acquire an imaginary part at larger values of $\epsilon$.
The appearance of a transition to complex dimensions is a recurrent aspect of melonic CFTs \cite{Giombi:2017dtl,Prakash:2017hwq,Giombi:2018qgp,Benedetti:2019eyl,Benedetti:2019rja}, and it is worth stressing that this is a very non-perturbative result, which would be hard to see from the perturbative series.

We notice also that at the merging point, the value $h_0=d/2$ means that the ``double-trace" quartic operator $(\phi^2)^2$ reaches marginality, because its dimension is $2h_0$, due to large-$N$ factorization.
The fact that fixed-point theories with cubic interactions could be destabilized by a quartic operator becoming relevant below some dimension (see for example \cite{deAlcantaraBonfim:1980pe} and references therein), and this happening above $d=4$ represents a case of dangerous irrelevant operator \cite{Amit:1982az}. In the case at hand we have a range $0\leq \eps \lesssim 0.264$ with real dimensions and $2h_{0}>d$, while for $\eps\gtrsim 0.264$ we have $2h_{0}=d+\im \a$, with $\a\in\mathbb{R}$; therefore, the double-trace operator never really becomes relevant, but it reaches marginality, with possible destabilizing effects (e.g.\ by leading to divergences in its conformal three-point functions \cite{Bzowski:2015pba}).
Moreover, a dimension of the form $d/2+\im \a$, as that of $h_0$, is expected to lead to an instability, because in the AdS/CFT picture it corresponds to bulk fields violating the Breitenlohner-Freedman bound \cite{Breitenlohner:1982jf}, and it has been conjectured to indeed signal a spontaneous symmetry breaking in \cite{Kim:2019upg}.

Following \cite{Giombi:2017dtl}, we can also compute the spectrum of the higher-spin operators characterized by spin $J\in\mathbb{N}$. For real fields, only even $J$ is in the spectrum, as the measure $\mu^{\D_\phi}_{h,J}$ in \eqref{eq:4pt}, written explicitly in \eqref{eq:measure}, vanishes for odd $J$.
The eigenvalues of the kernel are given by
\be 
k(h,J)=-2\frac{ \Gamma\left(1 - \f{1}{6}\eps\right) \Gamma\left(4 - \f{2}{3}\eps\right)  \Gamma\left(
   2 -\f{1}{3} \eps - \f{h - J}{2}\right) \Gamma\left(-1 + \f{1}{6}\eps + \f{h + J}{2}\right)}{ \Gamma\left(-1 + \f{1}{6}\eps\right)\Gamma\left(2 - \f{1}{3}\eps\right)  \Gamma\left(4 -\f{2}{3} \eps  -\f{h-J}{2} \right) \Gamma\left(
   1-\f{1}{6}\eps +\f{h + J}{2}\right)} \,,
\ee 
and denoting $h_{n,J}= 2\D_{\phi}^{(0)}+2n+J+z_{n,J}$, for the solutions of $k(h_{n,J},J)=1$ we obtain
\begin{align}
\begin{split}
     z_{0,2}=&-\eps\,,
\end{split}\\
\begin{split} \label{eq:z0J-SR}
     z_{0,J\geq 4}=&-\f{2\eps}{3}\left[1+\f{6\G(1+J)}{\G(3+J)}\right]+ \f{2\G(1+J)\eps^2}{9\G(3+J)}\bigg( 13 - 6 \g_E + 3 \psi(1 + J) - 9 \psi(3 + J) \bigg)\\&+\f{8\G(1+J)^2\eps^2}{\G(3+J)^2}\bigg( 1 + \psi(1 + J) - \psi(3 + J) \bigg)  +\cO(\eps^3)\,,
\end{split}\\
\begin{split}
    z_{1,J}=&-\f{2\eps}{3}\left[1 -\f{6\G(2+J)}{\G(4+J)}\right]+\f{2\G(2+J)\eps^2}{9\G(4+J)}\bigg(-19 + 6 \g_E - 3 \psi(2 + J) +9 \psi(4 + J)\bigg)\\&+\f{8\G(2+J)^2\eps^2}{\G(4+J)^2}\bigg(-1 + \psi(2 + J) - \psi(4 + J) \bigg)  +\cO(\eps^3)\,,
\end{split}\\
    z_{n,J}=&-\f{2\eps}{3}+\f{4\eps^2}{3n(n-1)}\f{\G(1+n+J)}{\G(3+n+J)} +\cO(\eps^3)\,,\quad (n\geq 2)\,,
\end{align}
with $\g_E$ standing for the Euler-Mascheroni constant and $\psi(z)$ is the digamma function.
We recognize that $h_{0,2}=d$, and the corresponding operator can then be identified with the energy-momentum tensor.

As well-known, for a unitary CFT, the conformal dimensions obey the lower bounds 
\be 
\label{eq:unitaryCFT}
h_{n,J}\geq 
\begin{cases} \f{d-2}{2} = 2-\f{\eps}{2} & \;\; \text{if } J=0\,, \\
d-2 +J = 4-\eps + J  & \;\; \text{if } J\geq 1\,.\end{cases}
\ee
The linear term in \eqref{eq:z0J-SR} would give a violation of the unitarity bound only for $J=1$, but since spin one is not in the spectrum, we conclude that there are no violations of unitarity at small $\eps$. We have numerically checked that the same conclusion holds all the way up to $\eps\sim 0.264$.

\paragraph{Long-range model ($\z=d/6$).}

Turning to the long range case, we recall the expression of the propagator 
\be 
G(p) = \f{\cZ_{LR}}{p^{d/3}}\,, \quad G(x) = \f{\Gamma(d/3)}{2^{d/3}\pi^{d/2}\G(d/6)}\f{\cZ_{LR}}{x^{2d/3}}\,,
\ee 
for which the same procedure as described previously leads to the following eigenvalues of the ladder kernel:
\be 
k_{d/6}(h,J)=2 g^2 \f{1}{(4\pi)^{d/2}} 
\f{\G(\f{d}{3}) \G\left(\f{d}{3}-\f{h-J}{2}\right) \G\left(\f{h+J}{2}-\f{d}{6}\right) }{ \G(\f{d}{6}) \G\left(\f{2d}{3}-\f{h-J}{2}\right) \G\left(\f{h+J}{2}+\f{d}{6}\right)}\,.
\ee 

In order to obtain the  solutions of $k_{d/6}(h,J)=1$, we do the following expansion. We write $h=h_{n,J}=2d/3+2n+J+z_{n,J}$, with anomalous dimension $z_n=\sum_{k>0} \a_k g^{2k}$,  and solve order by order in $g$ for arbitrary dimension $d<6$. Introducing the notation 
\be 
A_d = \f{\G(d/3)}{2^{d-2}\pi^{d/2}\G(d/6)}\,,
\ee 
we find the following solutions, at leading order in $g$:
\begin{itemize}
    \item for $d\neq 1,3$: 
    \be
    z_{n,J}=(-1)^{n+1}g^2A_d\f{\G(d/6+n+J)}{\G(d/2+n+J)\G(d/3-n)n!}\,,\quad (n,J\geq 0)\,,
    \ee 
    \item for $d=3$ there is a single solution, with $n=0$:
    \be 
    z_{0,J}=-\f{1}{2\pi^2} g^2\f{\G(1/2+J)}{\G(3/2+J)}\,,\quad (J\geq 0)\,;
    \ee 
    the absence of $n>0$ solutions is reminiscent of the $d=2$ case in Ref.~\cite{Benedetti:2019ikb}.
    \item for $d=1$:
    \be 
    z_{n,J}=(-1)^{n+1}g^2A_1\f{\G(1/3-n)\G(1/6+J+n)}{\G(1/2+J+n)n!}\,,\quad (n,J\geq 0)\,.
    \ee 
\end{itemize}

Demanding unitarity leads to a restriction on the allowed value of the coupling.
For $g=0$, we always have $h_{n,J}=2d/3+2n+J > d-2 + J$ for $d<6$ and $n\geq 0$, hence the free theory is unitary. As we turn on an infinitesimal $g^2>0$, and thus have $z_{n,J}\sim g^2$, the unitarity bounds can only be violated for $n=0$, $J>0$, and $d$ close to 6.
Indeed $z_{0,J}$ has a negative sign, and we see that at $d=6-\eps$, the following non-trivial bound on the coupling $g$ arises from the operators with $J>0$ and $n=0$: 
\be
g^2\leq \f{2^4\pi^3\Gamma(3+J)}{3\Gamma(1+J)}\eps\,.
\ee
The overall unitarity bound is given by the minimum of bounds over the admissible values of $J$, that is for $J=2$, corresponding to $g^2\leq 2^6 \pi^3 \epsilon$. 

\begin{figure}[htbp]
\centering
\includegraphics[height=5cm]{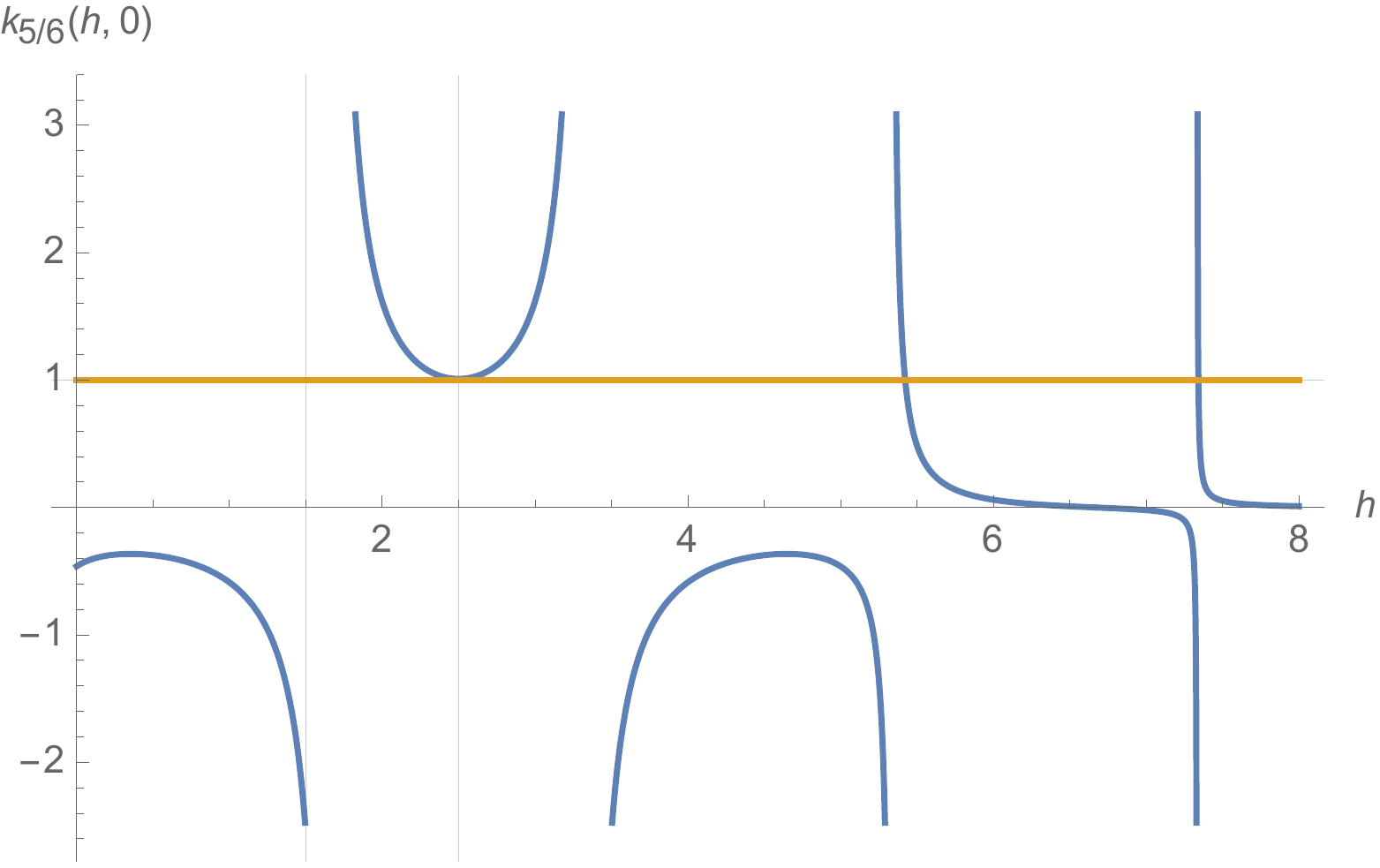}
\includegraphics[height=5cm]{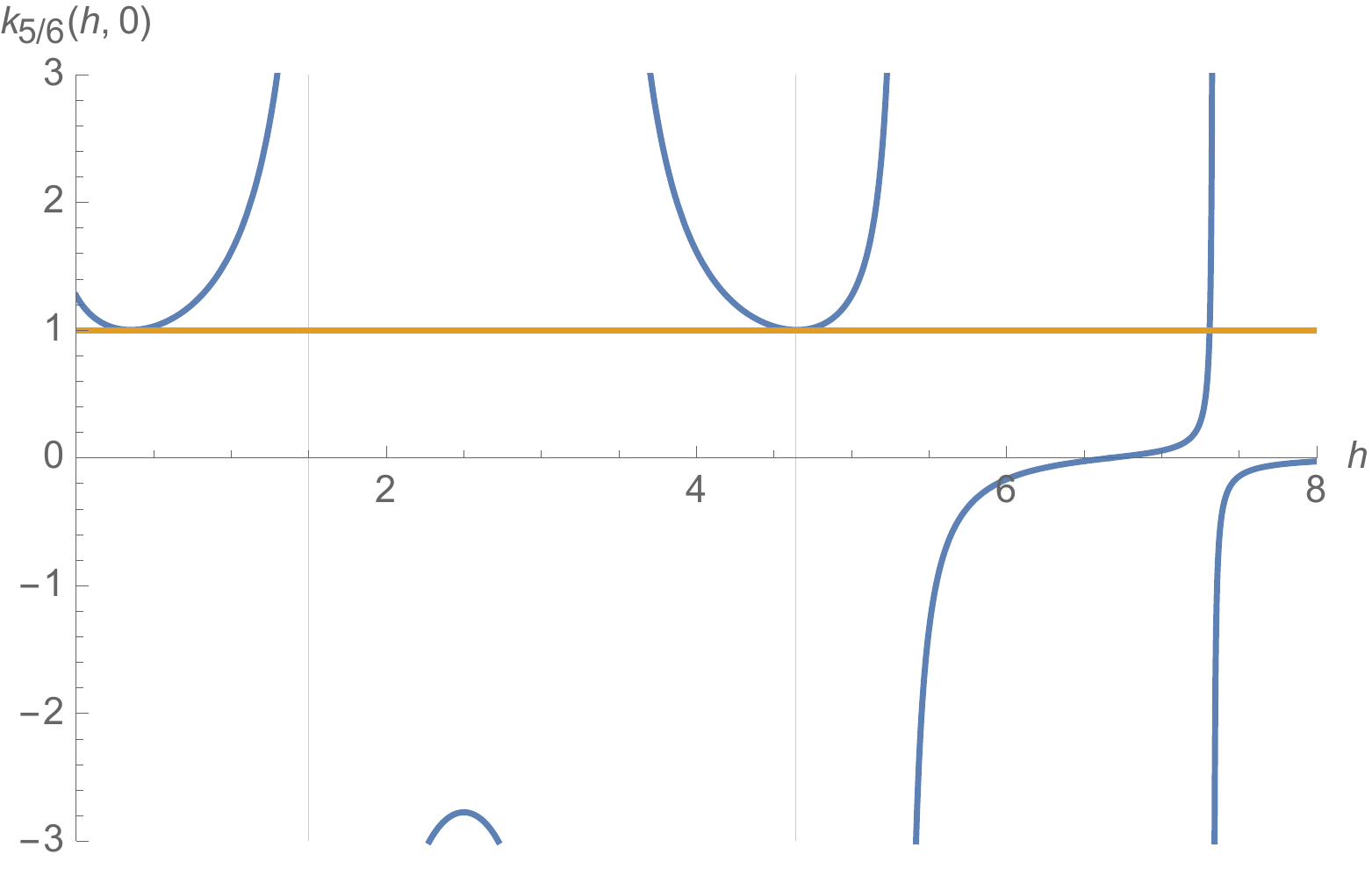}
\caption{Left: The eigenvalues $k_{5/6}(h,0)$ of the ladder-kernel in $d=5$, at the value $g=9.17$, when the lowest conformal dimensions $h_{0,0}$ merges with its shadow $h_{-1,0}$ at the value $5/2$. Right: The merging of the $h_{0,0}$ and $h_{1,0}$ conformal dimensions, at $g=\im 15.2$. In both plots, the leftmost vertical line (in light gray) represents the scalar unitarity bound $3/2$, while the rightmost one gives the value of $h$ at the crossing ($2.5$ and $4.64$).
}
\label{fig:unibound}
\end{figure}

\begin{figure}[htbp]
\centering
\includegraphics[height=5cm]{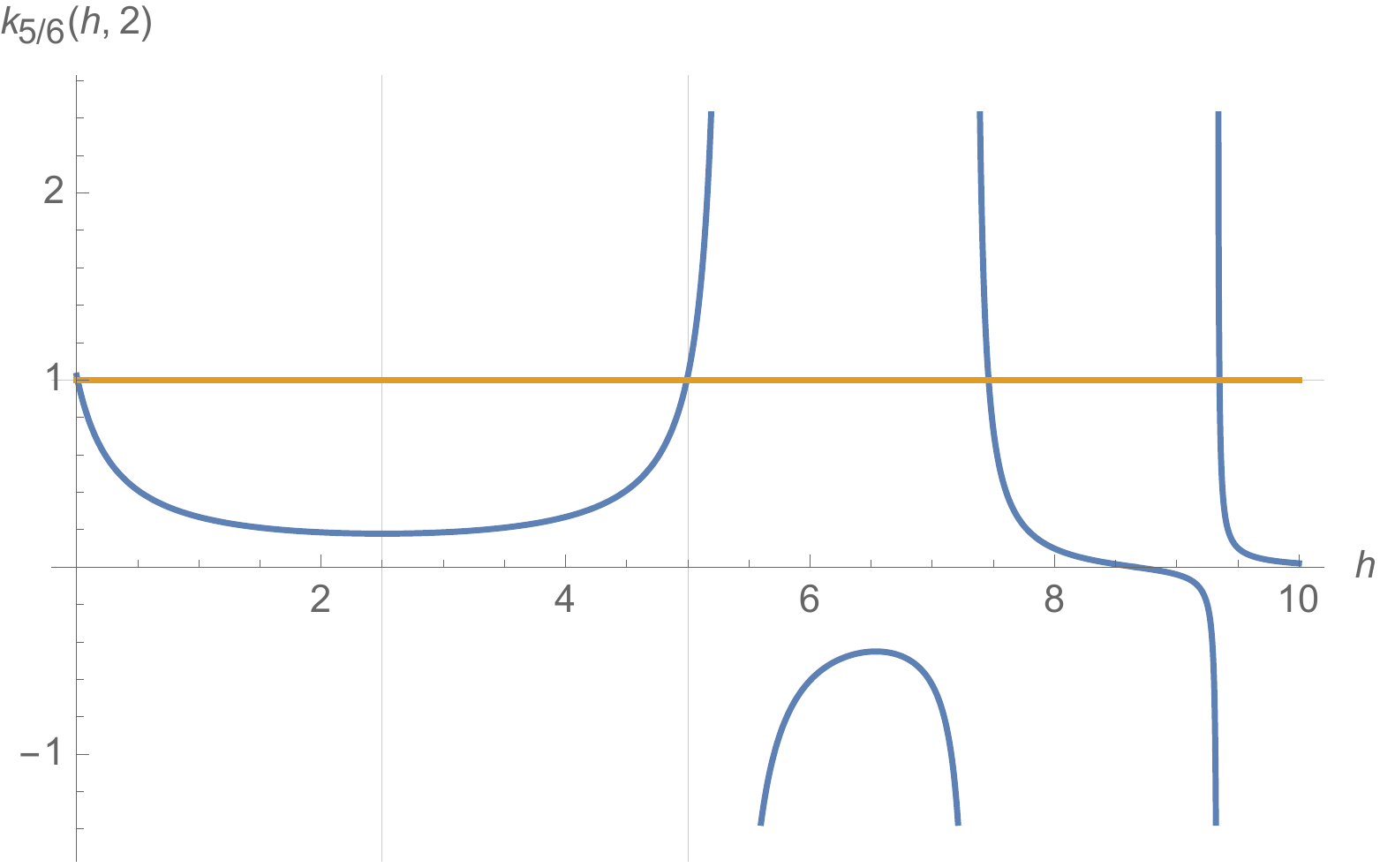}
\includegraphics[height=5cm]{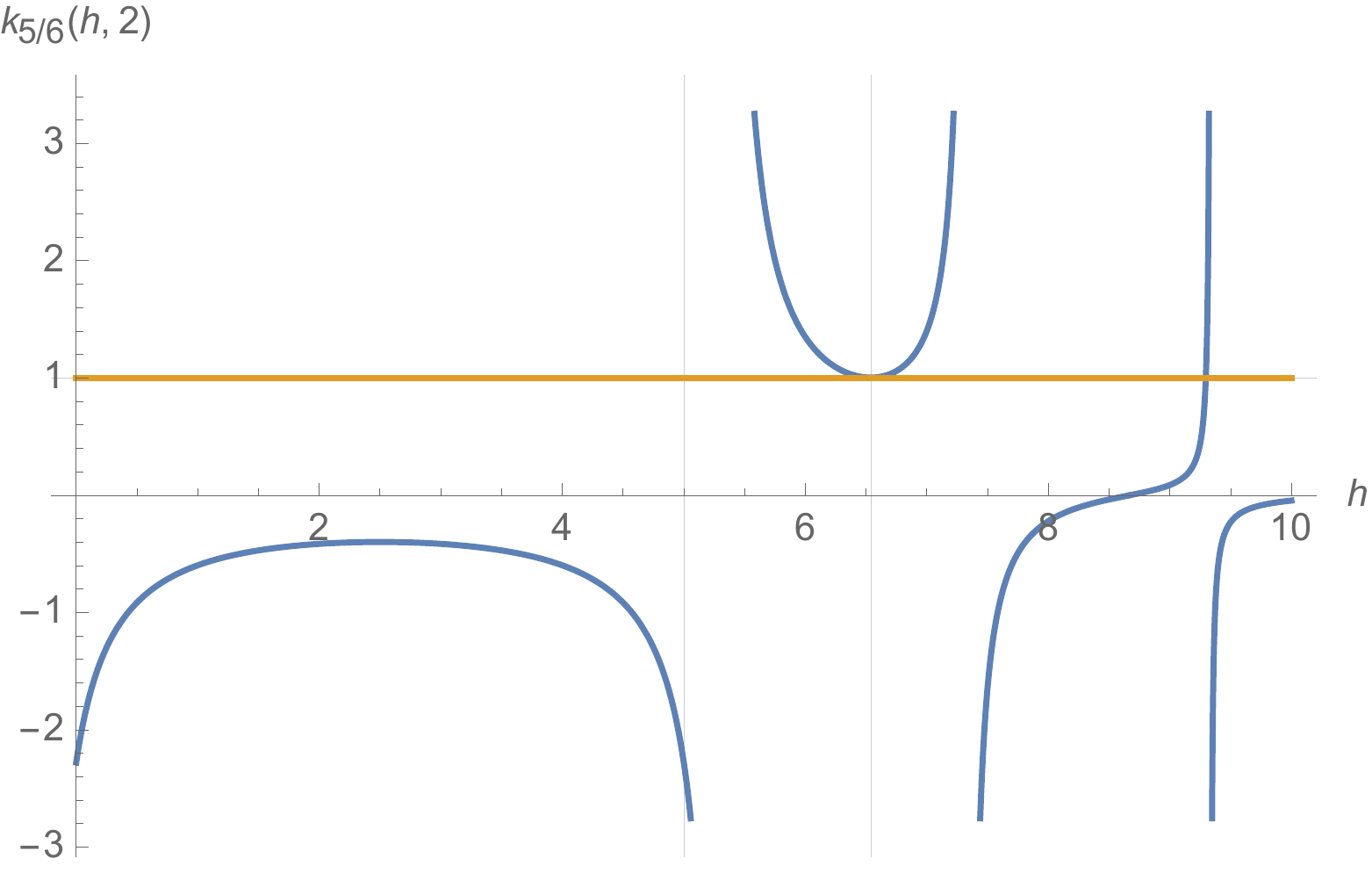}
\caption{Left: The eigenvalues $k_{5/6}(h,2)$ of the ladder kernel when the conformal dimension $h_{0,2}$ crosses the unitarity bound $h=5$, at $g=19.3$. It merges with its shadow at $h=5/2$, when $g=45.7$. Right: Merging of the conformal dimensions $h_{0,2}$ and $h_{1,2}$ at $h=6.54$, when $g=\im 28.8$. The vertical lines in light grey are again a guide for the eye, to show the unitarity bound and the abscissa of the merging.}
\label{fig:unibound2J}
\end{figure}

At larger values of $g$, and farther from six dimensions, in particular at integer dimensions, we have to check the unitarity bounds numerically.
We find that for any $d$ the appearance of a complex dimension, originating as in the short-range case from the merging of $h_{0,0}$ with its shadow $\tilde{h}_{0,0} = d-h_{0,0}$, occurs at a smaller value of $g$ than any possible unitarity violation.
Such merging is illustrated on the left panel of  Fig.~\ref{fig:unibound} for  $d=5$, at $g=  9.17$;
for $g\gtrsim 9.17$, the two solutions become complex, with real part equal to $d/2$.
On the left panel of Fig.~\ref{fig:unibound2J} we see instead the saturation of the $J=2$ unitarity bound taking place only at $g=  19.3$, hence there is no unitarity violation in the range of $g$ for which the CFT is real. 
In dimensions $d=4,\,3,\,2,\,1$ a similar situation is found, with the merging of $h_{0,0}$ with its shadow occurring at $g=3.69,\, 1.57,\, 0.72,\, 0.36$, respectively. Notice that these values remain below the critical coupling $g_{c,+}$ of eq.~\eqref{eq:g_+}, which for dimensions 5 to 1, have numerical values: 26.98, 16.49, 8.89, 4.65, 2.47.

For imaginary coupling, $g^2<0$ and $z_{0,J}$ is positive, hence the bounds \eqref{eq:unitaryCFT} are naturally obeyed for small $g$. 
At finite $g^2<0$, $h_{0,J}$ keeps growing, until it merges with $h_{1,J}$, as shown in the right panels of Fig.\ref{fig:unibound} and \ref{fig:unibound2J}. The merging for $J=0$ is the one that happens at the smallest value of $|g|$.

Notice that the merging of the two lowest dimensions, at $d=5$ and $g\simeq \im 15.2$, happens at the value $h_{0,0}=h_{1,0}\simeq 4.64$. Since at $g=0$ we have $h_{1,0}=16/3>5$, we see that the operator corresponding to $h_{1,0}$, which essentially is the $\phi\p^2\phi$ operator, crosses marginality (at $g=\im 13.5$) before the appearance of complex dimensions. Therefore, the fixed-point is probably destabilized by such operator, even before the merging occurs.
Interestingly, the $\phi\p^2\phi$ operator is the kinetic term of the short range model, and it is believed that its marginality crossing is responsible for the crossover from the long-range to the short-range Ising model  \cite{Sak:1973,Honkonen:1990,Behan:2017dwr,Behan:2017emf}; however, there are important differences to our situation: in the Ising case the crossover happens as one varies the value of the long-range exponent $\z$, and the $\phi^4$ interaction is not marginal in either version of the Ising model, while in the large-$N$ AR model we are varying the exactly marginal coupling of the cubic interaction; however, the coupling is never marginal in the short-range version of the AR model, hence there can be no continuous crossover from long-range to short-range AR models by simply varying $g$ in the former, at fixed $\z=d/6$. Moreover, as we have seen above, for $d<5.74$ in the spectrum of the short-range AR model there is an operator having complex dimension with real part equal to $d/2$, signaling an instability of the conformal phase of the model.

\section{Conclusions}
\label{sec:concl}

In this work, we have brought under a modern perspective the Amit-Roginsky model \cite{Amit:1979ev}, exploiting the techniques used recently within melonic conformal field theories. 
The model involves $N$ scalar fields forming an irreducible representation of $SO(3)$; invariance under such symmetry group allows for a unique cubic invariant interaction, built through the Wigner $3jm$ symbol, with $N=2j+1$. An appropriate rescaling of the coupling constant $\l$ to $\l\sqrt{N}$ then leads to a melonic limit at large $N$.
Such limit is an essential ingredient to the solvability of the SYK model, cornerstone of the $nAdS_2/nCFT_1$ holographic duality, or of tensor models, their counterparts without disorder. In this respect, the AR model ranges intermediately, not needing a disorder average, yet being simpler than tensors, for which a larger set of interactions needs to be taken into account. 

Introducing a fractional laplacian $(-\partial)^{2\z}$ in the kinetic term, we looked at $\z=1$ and $\z=d/6$ corresponding respectively to short- and long-range models in dimension $d$, following the footsteps of Refs.~\cite{Benedetti:2019eyl,Benedetti:2019rja} that dealt with quartic and sextic interactions. 
In the first case, we remained close to the upper critical dimension $d=6$ keeping the cubic interaction slightly relevant and using $\eps = 6-d$ to control our perturbative analysis. In the second case, for $0<d<6$, we tuned the dimension of the field to $\z=(d+\eps)/6$, before sending $\eps$ to zero. Also, since the propagator is non-local in this case, no wave-function renormalization is needed and the Schwinger-Dyson equations for the two-point function can be solved explicitly with a power-law ansatz $\cZ_{\rm LR}\,p^{-d/3}$, for a finite constant $\cZ_{\rm LR}$ that can be understood as the generating function of 3-Catalan numbers, providing a combinatorial resummation of the melonic diagrams.
Known properties of such function imply the existence of critical point for the effective coupling $g^2=\l^2 \cZ_{\rm LR}^3$, restricting it to a range $-\f12 g_{c,+}^2<g^2<g_{c,+}^2$, with $g_{c,+}$ given in \eqref{eq:g_+}.
Notice that we can allow the coupling to take imaginary value, which for an unbounded potential such as the cubic one is a rather sensible thing to do, as for example in the Lee-Yang model \cite{Fisher:1978pf,Cardy:1985yy}.
By contrast, the short-range version of the model requires a real coupling and a low-energy limit in order to lead to the same power-law solution, with a different proportionality constant $\cZ_{SR}$, as well as a wave-function renormalization.
We noticed that in the limit of large bare coupling, both constants $\cZ_{\rm LR}$ and $\cZ_{\rm SR}$ coincide.


A recurrent property of melonic theories is that at large $N$ the beta function of the coupling leading to the melonic limit does not contain vertex corrections, but only the wave-function renormalization. As a consequence, in the short-range version, for $d=6-\eps$, the two Wilson-Fisher real IR fixed points of \cite{Amit:1979ev} were recovered, while in the long-range one, we found a line of fixed points parametrized by the marginal coupling.

Moreover, we could use previous results on beta functions for generic cubic potentials, at finite $N$ and up to four loops \cite{Gracey:2015tta}, to discuss how $1/N$ corrections alter the large-$N$ results. The existence and nature of the fixed points are not changed in the short-range model (except at $N=13$, for which an imaginary coupling is needed instead). In the long-range case, since $1/N$ corrections break the marginality of the coupling, a more careful treatment is required in terms of a double-scaling parameter $\bar{\eps}=\eps\sqrt{N}$, in line with the works \cite{Fleming:2020qqx, Benedetti:2020sye}. 

At the fixed points, studying the conformal partial wave representation of the four-point function, we obtained the by now standard self-consistent equation for the conformal dimensions of the bilinear operators of arbitrary spin. At large-$N$ we find that in the short-range AR model the spectrum is real and above unitarity bounds up to $\epsilon\simeq 0.264$, when the smallest dimension of the scalar operators becomes complex, with real part equal to $d/2$, by merging with its shadow.
In the long-range case, we have the freedom to choose real or imaginary coupling, since it is exactly marginal.
In both cases we find that the spectrum is real and above unitarity bounds, for small $|g|$. As $|g|$ is increased, for real $g$ we find again a merging of the smallest scalar operator dimension with its shadow, while for imaginary coupling we find that it merges (for $d>3$) with the second smallest scalar operator dimension. Beyond the merging the respective dimensions become complex.
In the case of real $g$, like in the short-range case, the complex dimension has the form $h_{0,0}=d/2+\im \a$, with $\a\in\mathbb{R}$, which is expected to signal an instability, because in the AdS/CFT picture it corresponds to bulk fields violating the Breitenlohner-Freedman bound \cite{Breitenlohner:1982jf,Klebanov:1999tb}, and it has been conjectured to indeed signal a spontaneous symmetry breaking in \cite{Kim:2019upg}.
In the case of imaginary $g$, the complex dimensions have real part between $d/2$ and $d$, but the model is probably destabilized before reaching such merging, by the $\phi\p^2\phi$ operator crossing marginality.

We should notice that the melonic dominance at large-$N$ in the AR model has not been proved rigorously, but only based on a numerically checked conjecture. 
From a mathematical point of view it would be interesting to find a rigorous proof for the bound \eqref{eq:AR-bound}.

It is also tempting to think that due to the simplifications from the melonic limit, combined with the reduced complexity of the melonic two-point function in the case of cubic interaction, an all-order evaluation of the beta function or four-point function might be possible. The four-point function is given by the series of ladder diagrams of Fig.~\ref{fig:1overNladder}, decorated by melonic two-point functions. In the case of a standard propagator, and without melonic insertions, such ladder diagrams at arbitrary order have been computed explicitly in \cite{Usyukina:1993ch} in terms of polylogarithms. It would be worth to try to generalize such results to the case of the long-range propagator $p^{-d/3}$, or to try to obtain analogue results for the melonic two-point function diagrams.

\newpage

\appendix

\section{Conformal partial wave expansion for generalized free theories
with $\D_\phi = d/q$}
\label{app:CPW}

In this appendix we discuss, by means of the mean field theory example, a subtlety that can arise in the identification of the physical spectrum of the theory from the poles of the conformal partial wave expansion.
Our motivation for discussing this here is that in melonic CFTs with $q$-valent interaction one might be induced sometime to mistake a shadow operator for a physical one. For example, in the case of tensor models with sextic interactions ($q=6$) the appearance of a quartic operator in the OPE of two $\phi$'s has been erroneously reported in \cite{Giombi:2017dtl,Benedetti:2019rja}; in fact it can be checked that the supposed quartic operator dimension is actually that of the shadow of $\phi^2$, and that it differs from the dimensions of the possible quartic invariants computed from perturbation theory.\footnote{A similar correct identification of the extra pole with the shadow of $\phi^2$ has been noticed in another melonic CFT with sextic interaction, in \cite{Giombi:2018qgp}.}
The main observation does not rely on the presence of interactions, but only on the conformal dimension of $\phi$. The latter is fixed in the long-range models, hence we can take the non-interacting limit of a long-range theory with $\Delta_{\phi}=d/q$, and discuss the conformal partial wave expansion in such simplified setting.\footnote{For reference, we give here the expression of the ladder kernel for a long-range melonic theory with $q$-valent interaction, having $\z = \f{d(q-2)}{2q}$: 
\be 
k_{\z}(h,J)=(q-1) g^2 \f{1}{(4\pi)^{d(q-2)/2}}\left(\f{\G(\f{d}{q})}{\G(\f{d}{2}-\f{d}{q})}\right)^{q-2}
\f{ \G\left(\f{d}{q}-\f{h-J}{2}\right) \G\left(\f{h+J}{2}-\f{d(q-2)}{2q}\right) }{ \G\left(\f{d(q-1)}{q}-\f{h-J}{2}\right) \G\left(\f{d(q-2)}{2q}+\f{h+J}{2}\right)}\,.
\ee }

The four-point function in a generalized free CFT, also known as mean field theory, with a real scalar field of dimension $\Delta_{\phi}$ can be written as in Eq.~\eqref{eq:4pt} with vanishing four-point kernel:
\begin{equation} \label{eq:freeCPW}
\begin{split}
\cF(x_1,x_2,x_3,x_4)  & = \langle{\phi(x_1) \phi(x_3) \rangle \; \langle \phi(x_2) \phi(x_4)} \rangle 
 +  \langle{\phi(x_1) \phi(x_4) \rangle \; \langle \phi(x_2) \phi(x_3)} \rangle \\
&   =  \sum_J 
  \int_{\frac{d}{2}}^{\frac{d}{2}+\im\infty} \frac{dh}{2\pi \im}
   \; \r_{h,J}\,
     \Psi_{h,J}(x_i) \\
&   =  \sum_J 
  \int_{\frac{d}{2}-\im \infty}^{\frac{d}{2}+\im\infty} \frac{dh}{2\pi \im}
   \; \mu_{h,J}\,
     \cG_{h,J}(x_i)\,,
\end{split}
\end{equation}
where in the last step we used as standard (e.g.\ \cite{Liu:2018jhs}) the relation between conformal partial waves $\Psi_{h,J}$ and conformal blocks $\cG_{h,J}$:
\be
\Psi_{h,J}(x_i) =  \left(-\f12\right)^J  S_{\tilde{h},J}\, \cG_{h,J}(x_i)  +  \left(-\f12\right)^J  S_{h,J}\, \cG_{\tilde{h},J}(x_i) \,,
\ee
with
\be
S_{h,J} =\f{\pi^{d/2} \G(h-\f{d}{2}) \G(h+J-1) \G(\f{\tilde{h}+J}{2})^2}{\G(h-1) \G(d-h+J) \G(\f{h+J}{2})^2} \,.
\ee

The measure for the conformal block integral representation is given by:
\begin{align}\label{eq:measure}
\mu_{h,J} \, = & \, \left( \frac{ 1 + (-1)^J }{2} \right)
   \frac{  \Gamma(J+\frac{d}{2}) \Gamma(h-1)\Gamma(d-h+J)\Gamma(\frac{h +J}{2})^2
  }
  {  \Gamma(J+1) \Gamma(h-\frac{d}{2})\Gamma(h+J-1)\Gamma(\frac{d- h  +J}{2})^2 }
   \crcr
&  \quad \times  
\frac{  
 \Gamma( \frac{d}{2} - \Delta_{\phi})^2 
    \Gamma(\frac{ 2\Delta_{\phi} -d +h+J}{2}) 
    \Gamma(\frac{2\Delta_{\phi}-h+J}{2})
 }{  
 \Gamma( \Delta_{\phi})^2 
  \Gamma(\frac{2d-2\Delta_{\phi}-h+J}{2})\Gamma(\frac{d-2\Delta_{\phi} +h+J}{2}) 
 } \\
 = & \, \left(-\f12\right)^J \r_{h,J} S_{\tilde{h},J}\,.
 \nn
\end{align}

The conformal partial waves form a complete basis for field dimension in the principal series \cite{Simmons-Duffin:2017nub}, that is, for $\D_\phi =\f{d}{2}+ \im r $, with $r\in\mathbb{R}_{>0}$.
In such case, there are no additional (non-normalizable) contributions to \eqref{eq:freeCPW}, and in  the conformal block integral representation we can simply close the contour to the right to pick up the poles of the measure with ${\rm Re}(h)\geq d/2$. From these we should exclude the ``spurious'' poles of the measure that cancel with the poles of the conformal blocks \cite{Simmons-Duffin:2017nub}.
Such spurious poles are the poles of the $\Gamma(d-h+J)$ factor in the numerator of Eq.~\eqref{eq:measure}, which we will thus ignore.
For ${\rm Re}(h)>d/2$, the only possible poles come from $\Gamma(\frac{2\Delta_{\phi}-h+J}{2})$, i.e.:
\be \label{eq:free-h}
h_{n,J} = 2 \Delta_{\phi} + J + 2 n \,,  \qquad n\in \mathbb{N}_0 \,.
\ee
Notice that due to the combination $\Gamma(\frac{ 2\Delta_{\phi} -d +h+J}{2}) \Gamma(\frac{2\Delta_{\phi}-h+J}{2})$, if $\mu^{\Delta_{\phi}}_{h,J}$ has a pole at $h=h^*$, then it has a pole also at $h=d-h^*$, corresponding to the shadow operator. However, for $\D_\phi =\f{d}{2}+ \im r $ the shadows of \eqref{eq:free-h} have a negative real part, so they are not met when moving the contour in \eqref{eq:freeCPW} to the right.
We thus obtain
\be
\cF(x_1,x_2,x_3,x_4)  = \sum_{J,n} c^2_{h_{n,J}} \cG_{h_{n,J},J}(x_i) \,,
\ee
with squared OPE coefficients
\be
\begin{split}
c^2_{h_{n,J}} &= -\text{Res} \left[ \mu_{h,J}  \right]_{h=h_{n,J} }\\
& =  \f{(-1)^n}{n!}\left( 1 + (-1)^J \right)
   \frac{  \Gamma(J+\frac{d}{2}) \Gamma( \frac{d}{2} - \Delta_{\phi})^2 \Gamma( \Delta_{\phi} + J + n)^2
  }
  {  \Gamma(J+1) \Gamma( \Delta_{\phi})^2 \Gamma(\frac{d}{2}- \Delta_{\phi} - n)^2 } 
   \crcr
&  \quad \times    \frac{
    \Gamma( 2\Delta_{\phi}  +n+J-\f{d}{2}) 
    \Gamma(2 \Delta_{\phi} + J + 2 n-1)\Gamma(d-2 \Delta_{\phi} - 2 n) 
 }{ 
  \Gamma(d-2\Delta_{\phi}-n)\Gamma(\frac{d}{2} +n+J) 
  \Gamma(2 \Delta_{\phi} + J + 2 n-\frac{d}{2})\Gamma(2 \Delta_{\phi} + 2 J + 2 n-1)
 }  \,.
\end{split}
\ee
Notice that for real field dimension $\f{d}{2}-1<\D_\phi< \f{d}{2}$ (e.g.\ for $\D_\phi=d/q$ with $q>2$ and $d<d_c=\f{2q}{q-2}$), the sign factor $(-1)^n$ is canceled by the sign of the ratio $\Gamma(d-2 \Delta_{\phi} - 2 n)/\Gamma(d-2\Delta_{\phi}-n)$, and the OPE coefficients are therefore real in such case, as expected.

In order to understand which poles are physical in a theory with $\D_\phi=d/q$, we can keep $r>0$ while analytically continuing the real part to $d/q$ with $q>2$, and then send $r$ to zero.
All solutions \eqref{eq:free-h} with $J$ or $n$ greater than zero have real part greater than $d/2$ for $d<2q/(q-2)$, the latter being the dimension beyond which $\D_\phi$ violates the unitarity bound, hence we only need to worry about $h_{0,0}$.
As $q\to 4$, the solution $h_{0,0}$ and its shadow $\tilde{h}_{0,0}=d-h_{0,0}$ hit the line $d/2+\im \mathbb{R}$, and then for $q>4$ they swap place with respect to it.
The contour of integration should be deformed in such a way to keep  $h_{0,0}$ to its right and $\tilde{h}$ to its left.

As pointed out in \cite{Simmons-Duffin:2017nub}, the outcome of such contour deformation can equivalently be obtained from an undeformed contour along  the line $d/2+\im \mathbb{R}$, plus a non-normalizable contribution, by noticing that
\be
\begin{split}
\text{Res} \left[\r_{h,0} \Psi_{h,0}(x_i) \right]_{h=h_{0,0} }
& =  \text{Res} \left[ \mu_{h,0} \cG_{h,0}(x_i) \right]_{h=h_{0,0} } +
\text{Res}\left[\mu_{\tilde{h},0} \cG_{\tilde{h},0}(x_i) \right]_{h=h_{0,0} }\\
& =  \text{Res} \left[ \mu_{h,0} \cG_{h,0}(x_i) \right]_{h=h_{0,0} } -
\text{Res}\left[\mu_{h,0} \cG_{h,0}(x_i) \right]_{h=\tilde{h}_{0,0} }\,.
\end{split}
\ee
Due to the minus sign in the last expression, adding the conformal partial wave contribution to the conformal block integral representation with undeformed contour, and then moving such contour to the right of $\tilde{h}_{0,0}$, leads to an exact cancellation of the shadow contributions.

%

\bibliographystyle{JHEP}

\providecommand{\href}[2]{#2}\begingroup\raggedright\endgroup

\addcontentsline{toc}{section}{References}


\end{document}